\documentclass{aa}
\usepackage{xymtex}
\usepackage{newtxtext,newtxmath}
\usepackage{subcaption}
\usepackage{amsmath} 
\usepackage{xcolor}
\usepackage{float}
\usepackage{caption}
\usepackage{placeins}
\usepackage{amsmath}
\usepackage{lipsum}
\usepackage{multicol}
\usepackage{soul}
\usepackage[T1]{fontenc}
\usepackage{comment}
\usepackage{graphicx,tabularx,ragged2e,booktabs}
\usepackage[utf8]{inputenc}
\usepackage{hyperref} 
\UseRawInputEncoding

\def\refjnl#1{{\rm#1}}
\def\apjl{\refjnl{ApJ}}                
\def\apjs{\refjnl{ApJS}}               
\def\ao{\refjnl{Appl.~Opt.}}           
\def\aap{\refjnl{A\&A}}                

\def\be{\begin{equation}} 
\def\ee{\end{equation}}

\def\msun{{\msun}}

\def\gsim{\lower.5ex\hbox{\gtsima}} 
\def\lsim{\lower.5ex\hbox{\ltsima}} \def\gtsima{$\; \buildrel > \over 
\sim \;$} \def\ltsima{$\; \buildrel < \over \sim \;$} \def\prosima{$\; 
\buildrel \propto \over \sim \;$} \def\gsim{\lower.5ex\hbox{\gtsima}} 
\def\lsim{\lower.5ex\hbox{\ltsima}} 
\def\simgt{\lower.5ex\hbox{\gtsima}} 
\def\simlt{\lower.5ex\hbox{\ltsima}} 
\def\simpr{\lower.5ex\hbox{\prosima}}   
  
 \def\gtsima{$\; \buildrel > \over \sim \;$} 
\def\ltsima{$\; \buildrel < \over \sim \;$} 
\def\gsim{\lower.5ex\hbox{\gtsima}} 
\def\lsim{\lower.5ex\hbox{\ltsima}} 
\def\simgt{\lower.5ex\hbox{\gtsima}} 
\def\simlt{\lower.5ex\hbox{\ltsima}} 
\def\simpr{\lower.5ex\hbox{\prosima}}

\def\E3{{\cal E}_{\rm g}^{III}} 
\def\msun{\rm M_\odot}

\def\zsun{\rm Z_\odot}

\def\M*{M_*}

\def\Z*{Z_*}
\def\L*{L_*}

\makeatletter
\renewcommand*\aa@pageof{, page \thepage{} of \pageref*{LastPage}}
\makeatother

\begin{document} 
   \title{Black Hole merger rates in the first billion years in light of JWST data}

   \author{P. F. V. C{\'{a}}ceres-Burgos
          \inst{1} 
          \fnmsep
          \inst{2}
          \fnmsep
          \inst{3}
          \and P. Dayal
          \inst{1}
          \fnmsep
          \inst{4}
          \fnmsep
          \inst{5}
          \fnmsep
          \inst{6}
          \and P. Lira 
          \inst{2} 
          \fnmsep
          \inst{3}
          \and V. Mauerhofer
          \inst{1}
          \and F. P. Pratama
          \inst{7}
          \and M. Trebitsch
          \inst{8}
          }

   \institute{Kapteyn Astronomical Institute, University of Groningen, 9700 AV Groningen, The Netherlands\\
              \email{paula@astro.rug.nl}
              \and {Departamento de Astronom{\'{i}}a Cerro Calan, Universidad de Chile. Camino al observatorio 1515 Las Condes, Santiago de Chile}
              \and  {Millennium Nucleus on Transversal Research and Technology to Explore Supermassive Black Holes (TITANS)}
              \and {Canadian Institute for Theoretical Astrophysics, 60 St George St, University of Toronto, Toronto, ON M5S 3H8, Canada}
              \and {David A. Dunlap Department of Astronomy and Astrophysics, University of Toronto, 50 St George St, Toronto ON M5S 3H4, Canada}
              \and {Department of Physics, 60 St George St, University of Toronto, Toronto, ON M5S 3H8, Canada}
              \and {Departemen Fisika, FMIPA, Universitas Indonesia, Depok 16424, Indonesia}
              \and {LUX, Observatoire de Paris, Universit\`{e} PSL, Sorbonne Universit\`{e}, CNRS, 75014 Paris, France}
             }

  \abstract
   {Recent James Webb Space Telescope (JWST) discoveries have unveiled an abundance of faint and massive Active Galactic Nuclei (AGNs) at high redshifts (z=4-9), that surpass by 10 to 100 times the extrapolated bolometric (Bol) and ultraviolet (UV) luminosity functions (LF) from previous AGN campaigns. The two main models that are put forward to explain these observations correspond to light seeds ($\approx 150 M_{\odot}$) accreting in episodes of super Eddington, and heavy seeds ($\approx 10^3 - 10^5 M_{\odot}$) growing at the Eddington limit. Future gravitational observatories like the Laser Interferometer Satellite Antenna (LISA) will help disentangle these models by reporting the BH-BH merger events from mid to high redshifts.}
   {This work aims to report the predicted merger rates in the heavy seed scenario in light of recent JWST data. In our models we explore (i) instantaneous merging between BHs, (ii) delayed merging after a dynamical timescale, as well as extreme spin configurations (a=0.99, a=-0.99) to bracket BH mass growth.}
   {We use Delphi, a semi-analytical model that tracks baryonic physics over a hierarchical evolution of  dark matter halos through cosmic time within the first billion years of the Universe. We calibrate this model for it to simultaneously reproduce galaxy and JWST-AGNs observables.}
   {We show reasonable agreement with the Bolometric Luminosity function at $z=6$, where BHs must accrete 10 -100 times more gas than in previous works calibrated to pre-JWST data. However, we underpredict (overpredict) the bright end $10^{45.5}$erg s$^{-1}$ (all luminosity range) at z=7 (z=5) by 1$-$3.2 dex (0.22$-$1.6 dex). Regarding BH-BH merger events, the instantaneous (delayed) models predict a total of 28.06 (19.61) yr$^{-1}$ for BHs at z$\geq$5, which is within the range of merger rates reported in previous literature.}
   {}


   \maketitle

\section{Introduction}

\label{sec_intro}
The past years have seen an enormous increase in the number of active supermassive black holes (SMBHs; $ > 10^6 M_{\odot}$) detected in the early Universe at redshifts $z \gsim 4$. AGN samples collected using ground-based facilities \citep{Fan2006ConstrainingQuasars, Banados2016THEUNIVERSE, McGreer2013The82,Giallongo2015FaintUniverse,Niida2020TheSurvey} are now being supplemented by both photometrically \citep{Kokorev2024AFields, Akins2024COSMOS-Web:Assembly} and spectroscopically confirmed objects \citep{Harikane2023JWST/NIRSpecProperties, Ding2023Detection6, Yue2023EIGERzgtrsim6, Maiolino2023JADES.Mighty, Greene2023UNCOVERz5, Matthee2023LittleSurveys, Lin2024AWFSS} being detected with the James Webb Space Telescope (JWST) out to $z \sim 10.4$. AGNs selected with JWST instruments appear surprisingly abundant, reporting number densities that surpass by one to two orders of magnitude the extrapolated faint end of the predicted Bolometric and ultraviolet (UV) Luminosity Functions pre-JWST \citep{Shen2020The07, Niida2020TheSurvey}. Black Hole (BH) masses of JWST-AGNs, estimated to range between $10^6 - 10^9 M_{\odot}$ using single-epoch measurements, appear overmassive against the stellar masses of their galaxy host, deviating from predictions of local relations \citep{Reines2015RelationsUniverse}. Debate persists on whether scaling relations between BH and their galaxy properties evolve with redshift \citep{Pacucci2023JWSTModels} or arise from a selection bias effect that JWST instruments yield \citep{Li2024TipRelation}. Nonetheless, single-epoch techniques for BH mass tracers are probed to be robust in low redshift ($z<0.1$) AGNs \citep{Bentz2013THENUCLEI}, which also assume virial equilibrium in their Broad Line Region (BLR), a condition that is not clear to be satisfied for sources accreting close or above the Eddington limit \citep{Lupi2024SizeJWST,King2024TheQSOs}. Additionally, a subsample of JWST-AGNs, categorized as Little Red Dots (LRDs) for their red colours and compact morphology, showcase atypical SEDs, with uncertain origin in their UV continuum \citep{Akins2024COSMOS-Web:Assembly} and a lack of X-ray emission \citep{Ananna2024X-RayHoles, Maiolino2025iJWST/iSNe} 

Nevertheless, diverse theoretical models have been proposed to explain the origin and fast assembly of SMBHs. Among them are the formation of "heavy" BH seeds ($10^3 - 10^5 M_{\odot}$) accreting at the Eddington limit \citep{Kokorev2023UNCOVER:8.50,Bogdan2023EvidenceQuasar,Natarajan2024FirstCollapse,Dayal2024UNCOVERingEra} or through episodes of Super Eddington rates, or of light BH seeds ($\leq 10^2 M_{\odot}$) accreting at Super Eddington rates \citep{Lupi2016GrowingSeeds,Lupi2024SizeJWST,Regan2019Super-EddingtonHoles,Trinca2024EpisodicUniverse}. 

The study and detection of gravitational waves (GWs) from BH-BH mergers opens a new window to understanding the assembly of SMBHs through merger events. The collaboration from ground-based gravitational observatories LIGO-VIRGO-KAGRA have so far detected $\approx 90$ merging signals from stellar mass BHs ($\leq 10^2 M_{\odot}$) in the local Universe (\citet{Abbott2020GW190521:Mass,Abbott2021GWTC-2:Run,Abbott2023PopulationGWTC-3,Abbott2024GWTC-2.1:Run}, and others). Recent results from the data release 2 of the Pulsar Timing Array (PTA) \citep{Antoniadis2023TheUniverse}, which probes the nano-Hertz frequency of GWs using pulsars, have found a large amplitude signal suggestive of a cosmic GW background from SMBH binaries above $10^7 M_{\odot}$ (SMBHBs) at low redshifts $z<1$. Future gravitational wave observatories will aim to detect BH mergers in the intermediate mass regimes $10^3 - 10^7 M_{\odot}$ at higher redshifts $z\geq2$. Among these are the Laser Interferometer Satellite Antenna (LISA) \citep{Amaro-Seoane2023AstrophysicsAntenna}, which would detect mergers from $10^3 - 10^7 M_{\odot}$ BHs in their last stages of inspiraling and subsequent merger up to $z \sim 20$, and the Einstein Telescope (ET) sensitive to detect $\leq 10^3 M_{\odot}$ BH binary systems at $z \geq 2$ \citep{Punturo2010TheObservatory}. Detecting signals of massive BH mergers through cosmic time will give us valuable insight into the growth history of BHs via mergers, and an even stronger constraining power on the theoretical models of BH formation and growth \citep{Valiante2018ChasingStatistics, Amaro-Seoane2023AstrophysicsAntenna} than what we can currently speculate from their electromagnetic emission. 

So far, many theoretical predictions of BH merger rates have been drawn using analytical, semi-analytical and cosmological hydrodynamical simulations. For the heavy seed scenario specifically, they range between 0.02 to 244 $yr^{-1}$ for $z>5$ \citep{Dayal2019TheAstronomy, Sesana2007TheStream,Sesana2011ReconstructingWaves,Ricarte2018TheSeeds,Barausse2020MassiveFeedback, Klein2016ScienceBinaries, Katz2020ProbingLISA, Bonetti2019Post-NewtonianLISA}. In contrast, predictions for light seeds are usually larger, given their more abundant number densities as remnants of large pop III stars, with merger rates that range between 26.11 to 140.45 yr$^{-1}$ for $z>5$ \citep{Dayal2019TheAstronomy,Ricarte2018TheSeeds,Barausse2020MassiveFeedback,Klein2016ScienceBinaries,Sesana2007TheStream,Sesana2011ReconstructingWaves}. 
However, most of these predictions are based on models calibrated on pre-JWST data. Therefore, we find it relevant to revisit these predictions using a semi-analytical model calibrated against JWST observations for AGNs and galaxies. 
We use The SAM code{ \sc Delphi}, that tracks the hierarchical assembly of dark matter halos and their baryonic components (including stars and black holes) between $z \sim 26-4.5$.
The advantage of semi-analytical models (SAMs) lies in their computationally cheap calculations and their capability to track galaxy and BH growth, exploring a wide parameter space. Whereas hydrodynamical cosmological simulations are restricted to seed massive BHs $\approx 10^{4-6} M_{\odot}$ \citep{Salcido2016MusicSimulations,Tremmel2018DancingMergers,Katz2020ProbingLISA,Volonteri2020BlackSimulations}, limited to the resolution of their dark matter particles ($\approx 10^5 M_{\odot}$).

We adopt a $\Lambda$CDM model, and consider the Planck 2018 results for the cosmological constants \citep{Aghanim2020PlanckParameters} ($\Omega_{\Lambda}$, $\Omega_{m}$, $\Omega_{b}$, $h$, $\sigma_{8}$) = (0.6889, 0.3097, 0.04897, 0.677, 0.811), where $\Omega_{\Lambda}$ is the dark energy density parameter, $\Omega_{m}$ the total matter density (including dark and baryonic matter), $\Omega_{b}$ the baryonic matter density, $h$ the dimensionless reduced Hubble constant ($H_0 = 100h$km$\ $s$^{-1}$Mpc$^{-1}$), and $\sigma_{8}$ the matter fluctuation amplitude. We use a Kroupa initial mass function \citep[IMF;][]{Kroupa2001OnFunction} between $0.1-100 \msun$, with a slope of -1.3 (-2.3) between 0.1-0.5 (0.5-100) $M_{\odot}$,  and a mass weighted solar metallicity value of $\zsun = 0.0122$ \citep{Asplund2005TheComposition}. Finally, we express all magnitudes in the standard AB system \citep{Oke1983SecondarySpectrophotometry}.

This work is structured as follows: In section \ref{theoretical_framework} we explain the theoretical framework and the models explored in this work. In section \ref{sec:calibrate}, we show how our models compare to AGN observables. In section \ref{sec:emergence_bhs} we describe the emergence of the stellar to BH mass relation and BH accretion density implied from our models, along the merger rates implied for each model. Finally, in section \ref{sec_conclusions} we discuss our results and conclusions.

\section{The Theoretical model}

\label{theoretical_framework}
For this work we use {\sc Delphi} (Dark matter and the Emergence of gaLaxies in the ePocH of reIonization), a semi-analytical model that tracks galaxy and black hole formation between $z \sim 40-4.5$. First introduced in \citet{Dayal2014EssentialFormation}, since then has undergone many modifications to explore predictions for gravitational waves from BH mergers \citep{Dayal2019TheAstronomy} in the first billion years of cosmic time, the physical properties and growth of BHs in high-$z$ galaxies \citep{Piana2020TheHoles, Piana2021TheGalaxies}, metal and dust evolution in high-$z$ galaxies \citep{Dayal2022,Mauerhofer2023TheEra, Mauerhofer2025SynergisingYears}, as well as the impact on reionization from BHs in light of JWST data \citep{Dayal2024UNCOVERingEra}. From now onwards we will refer to \citet{Dayal2019EarlyEffects} as D19.

We build our theoretical framework for gas, stellar, BHs, dust and metal evolution as in \citet{Dayal2014EssentialFormation, Dayal2022, Dayal2024UNCOVERingEra}, but including the recent modifications proposed in \citet{Mauerhofer2025SynergisingYears}, which adopts cold gas fractions and star formation efficiencies drawn from the {\sc Sphinx}$^{20}$ simulation \citep{Rosdahl2022LyCReionization}. 

\subsection{Dark matter merger trees and gas assembly}

The dark matter halo merger tree was constructed following the \citet{Parkinson2007GeneratingTrees} algorithm, which describes a hierarchical assembly of halos based on the Extended Press-Schechter theory. 
 The merger tree used in this work has halos with masses ranging between $\log(M_{h}/M_{\odot}) = 8 - 14$ starting at $z=4.5$, with 100 halos per dex distributed uniformly in log-space, i.e., 600 trees in total. This merger tree evolves backwards up to redshift 26, in fixed timesteps of $\Delta t=30$ Myr. At each timestep, a halo of mass $M_1$ can either fragment into 2 halos of mass $M_2$, where $M_{res} < M_2 < M_1/2$, or lose a fraction of its mass below the resolution limit ($M_{res}$). The lost fraction of dark matter accounts for the contribution of unresolved halos i.e. smooth-accretion of dark matter ($M_{\rm dm}^{\rm sa}$) from the intergalactic medium (IGM). The mass resolution of this merger tree is $10^8 M_{\odot}$. Each halo at $z = 4.5$ is assigned a comoving number density $dN/dlogM_h [{\rm cMpc^{-3} dex^{-1}}]$ by comparing to the Sheth-Tormen halo mass function (HMF) \citep{Sheth1999Large-scaleSplit} at that redshift. This number density is inherited by all the descendants of the corresponding halo. 
 
 In the starting leaves of our merger tree (first-generation halos resolved in the SAM), we begin with an initial gas mass equal to the cosmological ratio between baryons and dark matter as $M_g^i = (\Omega_b/ \Omega_m) M_h$. Subsequently, the total gas mass inhereted by the descendant halo is computed as the leftover gas after stellar and BH processes accompanied by a factor of smoothly accreted gas from the IGM ($M_{g}^{\rm sa}$). We find $M_{g}^{\rm sa}$ by multiplying the smoothly accreted dark matter mass ($M_{\rm dm}^{\rm sa}$) with the cosmological ratio $(\Omega_b/ \Omega_m)$.

\subsection{Star formation, feedback and UV luminosity}

The stellar mass produced within each timestep is computed as the effective stellar efficiency ($f_{\star}$) times the initial gas mas, i.e. $M_{\star} = f_{\star}^{\rm eff} M_g^i$, where $f_{\star}^{\rm eff} = min[f_{\rm cold} \cdot f_{\star}^{\rm sphinx}, f_{\star}^{\rm ej}]$. The $f_{\rm cold}$ and $f_{\star}^{\rm sphinx}$ are the cold gas fraction and the star formation efficiency retrieved from {\sc Sphinx}$^{20}$, respectively. Both quantities depend on the mass and redshift of the halo, and are assigned accordingly. More details on retrieval and tabulation of these values from {\sc Sphinx}$^{20}$ simulations can be found in \citet{Mauerhofer2025SynergisingYears}. The term $f_{\star}^{\rm ej}$, on the other hand, is the star formation efficiency necessary to eject all remaining gas from the halo by type II SNe (SNII) feedback:

\begin{equation}
    f_{\star}^{\rm ej} = \frac{v_c^2}{v_c^2 + f_w^{\star} \nu E_{51}}
\end{equation}

where $v_c$ is the halo rotational velocity, $\nu$ is the supernova rate ($1.113 \times 10^{-2} M _{\odot}^{-1}$ for a Kroupa IMF), and $f_w^{\star}$ is a free parameter that sets the fraction of SNII energy fraction coupling to gas. Halos where $f_{\rm cold} \cdot f_{\star}^{\rm sphinx} > f_{\star}^{\rm ej} $ are SNII feedback limited, i.e., all their gas gets ejected after SNII; otherwise, they are efficient star formers. 

For each galaxy, the intrinsic UV luminosity between 1450-1550 \r{A} ($L_{\rm UV}^{\rm int}$) comes from the SED calculated by convolving its star formation history (assuming constant star formation for the stellar mass formed within the 30 Myr timestep) using Starburst99 (SB99) stellar population synthesis code \citep{Leitherer1999STARBURST99FORMATION}. We adopt the Geneva High mass loss tracks \citep{Meynet1994Grids0.040}, using metallicities of 0.001, 0.004, 0.008 and 0.02 $Z_{\odot}$. For each galaxy, we interpolate between these values based on its gas-phase metallicity.
 The observed UV luminosity ($L_{\rm UV}^{\rm obs}$) is estimated assuming an internal extinction, corresponding to a homogeneous mixture of stars and dust: $L_{\rm UV}^{\rm obs} = L_{\rm UV}^{\rm int} \frac{1 - e^{-\tau_{\rm d}}}{\tau_{\rm d}}$, where $\tau_{\rm d}$ is the dust optical depth for a given galaxy (explained in Sec. \ref{subsec:dust}). 

\subsection{BH seeding, accretion, feedback and luminosities}

Heavy BHs are seeded in all starting metal-free halos down to redshift 12.8. We randomly draw the initial BH masses from a log-uniform distribution in the $10^3 - 10^5 M_{\odot}$ mass range. Once seeded, BHs accrete the minimum gas mass between an Eddington fraction ($f_{\rm Edd}$) or a fraction of the remaining gas after SNII  feedback ($f_{\rm acc}^{\rm BH}$). Hence, the accreted mass is determined as: 
\begin{align}
    M_{\rm BH}^{\rm acc} = min[(1-\epsilon_r(a))\, f_{\rm Edd}\, \dot{M}_{\rm Edd}\, \Delta t, \\ (1-\epsilon_r(a))\, f_{\rm acc}^{\rm BH}\, M_{g}^{\star}] \nonumber 
\end{align} 

where $\dot{M}_{\rm Edd}$ is the Eddington mass accretion of the BH ($\dot{M}_{\rm Edd} = 4 \pi G m_p\ (\epsilon_r(a)\ c\ \sigma_{T})^{-1}\ M_{BH} $) and $\epsilon_r(a)$ the BH radiative efficiency for a given dimensionless spin $a$ (details in Sec. \ref{sec:bh_spins}). We decide not to split the gas into cold and hot components for BH accretion as we do in star-forming processes, since {\sc Sphinx}$^{20}$ does not include AGN physics, and letting BHs accrete from a given component would give rise to inconsistencies with the cold gas fraction sampled from {\sc Sphinx}$^{20}$. 
Both $f_{\rm Edd}$ and $f_{\rm acc}^{\rm BH}$ are free parameters in our models that are implemented with a Gaussian scatter in log-space with a 0.5 dex standard deviation. These parameters can adopt two different values depending on whether the halo mass surpasses the critical threshold $M_{h}^{\rm crit}=10^{11.25}[\Omega_m (1+z)^3 + \Omega_{\Lambda}]^{-0.125} (M_{\odot})$ \citep{Bower2017TheEnd}.  In our framework, the accretion parameters are largest when $M_h\geq M_h^{crit}$. The physical reasoning behind $M_h^{\rm crit}$ is that for lower mass halos, star formation-driven outflows would prevent the build-up of gas in the galaxy's center, inhibiting the growth of the nucleated BH. To reproduce the dispersion of the BH to stellar mass relation found in observations, we apply a gaussian scatter of 0.5 dex in log-space for the $M_h^{\rm crit}$ threshold that each halo needs to surpass. We chose this magnitude of dispersion (0.5 dex) because it is typically used in analytical models that reproduce early galaxy observables \citep{Nikopoulos2024AUniverse}.

We define the effective Eddington fraction ($f_{\rm Edd}^{\rm eff}$) as the ratio between the accreted gas mass ($M_{\rm BH}^{\rm acc}$) and its Eddington mass accretion ($(1-\epsilon_r(a))\, \dot{M}_{\rm Edd}\, \Delta t$) within the timestep. 

The strength of BH feedback, i.e., the fraction of energy from BH accretion coupled to winds, is regulated by a free parameter ($f_w^{BH}$), which affects the gas availability in halos. The bolometric luminosity of BHs with $f_{\rm Edd}^{\rm eff} \geq 1 \%$ is calculated as $L_{\rm Bol} = \epsilon_r(a)\, M_{\rm acc}^{\rm BH}\, c^2/ (\Delta t \times (1-\epsilon_r(a)))$, 
where $c$ is the speed of light. This luminosity decays instantaneously in the next $\Delta t$. For lower effective Eddington fractions ($f_{\rm Edd}^{\rm eff}<1\%$), models predict a transition from a UV bright, thin disk to a radiatively inefficient, optically thin accretion flow, where the bulk of the thermal energy is carried into the BH rather than being radiated (\citet{Narayan1994Advection-dominatedSolution}, \citet{Yuan2014HotHoles}), therefore their SED changes drastically, with a decrease in the UV/Optical regime \citep{Esin1997ADVECTION-DOMINATED1991, Chiaberge2006LowNGC4565, Younes2012StudyCorrelations, Almeida2018TheProperties}. For this regime of inefficient accretion, we consider BHs to be non-luminous. We find that neglecting inefficiently accreting BHs ($f_{\rm Edd}^{\rm eff}<1 \%$) does not impact our results.

\subsection{Dust, metal evolution and opacities}
\label{subsec:dust}

In our model, the interlinked evolution of metals and dust are followed using the prescriptions detailed in \citet{Dayal2022} and \citet{Mauerhofer2023TheEra}. In brief, we employ a set of coupled differential equations that describe the production, ejection, and astration of metals and dust, including the growth of dust grains and their destruction into metals via SNII shocks; we assume perfect mixing of metals, dust and gas. At early times ($z>4.5$), SNII are the primary drivers of dust production (producing 0.5$M_{\odot}$ of dust per SNII); AGB (asymptotic giant branch) stars contribute negligibly \citep{Dayal2010DetectingSubmillimetre,Lesniewska2019Dust6-8.3}. For metal production, we consider the contribution of all SN and AGB stars between 2-50 $M_{\odot}$, and use the metal yields results from \citet{Kobayashi2020TheUranium}. We employ this mass constraint because stars below 2$M_{\odot}$ end their life at longer timescales than the age of the Universe at $z>4.5$, while stars $>50 M_{\odot}$ are assumed to collapse into black holes without any metal yield. With the incorporation of cold gas fractions, we calculate the destruction and growth of dust grains in the warm and cold ISM, respectively, as detailed in \citet{Mauerhofer2023TheEra}.

Subsequently, we estimate UV opacity for both star formation and black holes.
Since dust and gas are perfectly mixed, we equate their distribution to a sphere of radius $r_{\rm gas} = r_{\rm dust}$. Results from the Atacama Large millimetre Array (ALMA) Rebels survey find a negligible evolution of the effective radii of [CII] 158 $\mu$m at a fixed UV luminosity in the 4 to 7 redshift range \citep{Fudamoto2022The4}. We incorporate this information \citep[tested in][]{Mauerhofer2023TheEra} and calculate the evolution of the gas radius as $r_{\rm gas} = 4.5\, \lambda\, \frac{1+z}{6}\, r_{\rm vir}$, where $\lambda$ is the halo spin parameter with an average value of $\lambda = 0.04$ \citep{Dayal2019EarlyEffects}, and $r_{\rm vir}$ is the virial radius of the halo. The UV optical depth for star formation is calculated as $\tau_{\rm gal}=3\, M_{\rm dust} [4 \pi\, a\, s\, r_{\rm dust}^2]^{-1}$, where $M_{\rm dust}$ is the total dust mass, and $a=0.05\ \mu$m and $s=2.25$ gm cm$^{-3}$ correspond to the size and density of a carbonaceous/graphite dust grains, respectively \citep{Todini2001DustSupernovae, Nozawa2003DustSupernovae}.

\subsection{BH spins}
\label{sec:bh_spins}

In this work, we explore the maximal spin scenarios for all BHs to bracket BH mass growth and understand its implications on galaxy and AGN observables. We use the dimensionless definition of the spin $a = J_{\rm BH}/J_{\rm max}$, where $J_{\rm BH}$ is the angular momentum of the BH and $J_{\rm max} = GM_{\rm BH}^2/c$ its the maximum angular momentum. If $a>0$, we consider the BH to co-rotate with the accreting material, whereas $a<0$ corresponds to the counter-rotating case. The radiative efficiency ($\epsilon_r$) is defined as a function of the spin \citep{Bardeen1972ROTATINGRADIATION}. For this work, we consider three spins for all BHs: Extreme prograde/co-rotating scenario $a=0.99$ that yields $\epsilon_r = 0.263$, extreme retrograde/counter-rotating scenario a=-0.99 that yields $\epsilon_r = 0.037$, and the "canonical" spin $a=0.67$ that yields the radiative efficiency $\epsilon_r \approx 0.1$ most commonly assumed in observational work \citep[e.g.,][]{Ueda2014TowardPopulations}. Since BH accretion ($M_{\rm BH}^{\rm acc}$) scales as $(1-\epsilon_r(a))$ or $(1 - \epsilon_r(a))/ \epsilon_r(a)$ depending on the accretion mode, the extreme prograde and retrograde scenario will minimize and maximize the accreted mass, respectively. On the other hand, the bolometric luminosity scales as $\epsilon_r(a)/(1 - \epsilon_r(a))$ times $M_{\rm BH}^{\rm acc}$, therefore it is only on the mode of sub-Eddington accretion that the luminosity scales with $\epsilon_r(a)$.

\subsection{BH and galaxy mergers}
\label{sec:mergers}

Whenever two or more dark matter halos with galaxies and BHs merge, we consider two plausible scenarios: (1) Instantaneous merging: DM halos, galaxies and BHs merge instantaneously, and (2) Delayed merging: galaxies and BHs merge after a dynamical friction timescale \citep{Lacey1993MergerFormation}:

\begin{align}
    \tau_{\rm DF} = f_{\rm df}\, \Theta_{\rm orbit}\, \tau_{\rm dyn}\, \frac{M_{h}}{M_{\rm sat}} \frac{0.3722}{\ln(M_{h}/ M_{\rm sat})}.
    \label{dynamical_friction_time}
\end{align}

Here, $M_{h}$ is the mass of the main halo including all satellites, $M_{\rm sat}$ is the halo mass of the merging satellite, $\tau_{\rm dyn}$ is the dynamical timescale that can be expressed as the half period of a circular orbit for the virial radius of the halo ($\pi\, R_{\rm vir\rm }/V_{\rm vir} = 0.1\, \pi\, t_{H}(z) $), $f_{\rm df}$ is the efficiency of tidal stripping that we impose to be $f_{\rm df}=1$, and $\Theta_{\rm orbit}$ is the angular momentum of the satellite halo normalized to a circularized orbit. From \citet{Cole2000HierarchicalFormation} it is found that $\Theta_{\rm orbit}$ follows a lognormal distribution as $\ln\, \Theta_{\rm orbit} = -0.14 \pm 0.26$, hence we randomly draw values for each merger assuming this distribution. In our prescription for delayed merging, satellite galaxies that are orbiting the main system will cease star formation, and their BH (if present) will stop accreting gas.

The dynamical friction timescale $\tau_{\rm DF}$\ must be taken as a lower (i.e. optimistic) limit on the actual time it takes BHs to merge. This is because dynamical friction becomes inefficient at separations of a few kpc and below \citep{Begelman1980MassiveNuclei}. Further sinking from a few to one kpc scales among the interacting BHs becomes a stochastic process, highly sensitive to the properties of the merging galaxies. These include their stellar mass ratios, stellar masses, morphologies, \citep{Tremmel2018DancingMergers, Biava2019TheModels, Bortolas2021TheBinaries}, and even the masses of the participating BHs \citep{Pfister2019TheGalaxies}. At separations below $< 100$ pc scales, orbital decay is driven by a combination of processes such as dynamical friction against the stellar and/or gaseous density profiles, stellar hardening, gas migration, or third-body interactions \citep{Escala2005THEDISK, Haiman2009TheNuclei, Colpi2014MassiveCoalescence, Dosopoulou2017DynamicalProblem, Bonetti2019Post-NewtonianLISA}. Finally, at sub-pc distances (typically $\approx 10^{-4}-10^{-3}$pc for equal mass binaries of $10^{6} - 10^7 M_{\odot}$), gravitational wave (GW) driven inspiral dominates and ultimately leads to coalescence \citep{Amaro-Seoane2023AstrophysicsAntenna}. Naturally, accounting for all these sub-kpc dynamics would increase the delays of BH mergers, and reduce the reported merger rates found in this work, an effect that has been demonstrated in other SAMs works \citep{Bonetti2019Post-NewtonianLISA, Barausse2020MassiveFeedback}.

Moreover, since BHs in our prescription merge instantaneously or after $\tau_{DF}$, we do not account for the possibility that they become wanderers, BHs that never merge with the main BH. A growing number of studies report that wandering BHs can represent a significant fraction of the total BH population \citep{Ricarte2018TheSeeds, Tremmel2018DancingMergers, Izquierdo-Villalba2020FromEnvironments, Bellovary2021TheGalaxies, DiMatteo2022ANoon, Chen2022DynamicalPredictions}. 
For now, we limit this work to the most optimistic scenario for BH mergers, in which mergers occur as efficiently as our dark matter merger tree resolution allows. This assumption maximises the predicted BH merger rate and facilitates the early growth of massive BHs, as suggested by recent JWST results. Nonetheless, incorporating more realistic sub-kpc timescales remains a key goal for future iterations of this model.

We consider the fiducial model as the one that assumes instantaneous merging of galaxies and BHs and a canonical radiative efficiency $\epsilon=0.1$ that yields a spin of $a=0.67$ (see Section \ref{sec:bh_spins}). A summary of the models explored in this work can be found in Table \ref{models}.

\begin{table}[]
\small
\caption{Models explored in this work.}
\centering

\begin{tabular}{llll}

\hline
Name model & Merging timescale  & spin (a) & $\epsilon_r$ \\ \hline
ins (fiducial)      & 0                  & 0.67      & 0.99        \\
del       & $\tau_{DF}$ (Eq. 
 \ref{dynamical_friction_time}) & 0.67      & 0.99        \\
ins-sp099   & 0                  & 0.99        & 0.263        \\
del-sp099  & $\tau_{DF}$ (Eq. 
 \ref{dynamical_friction_time}) & 0.99        & 0.263        \\
ins-sn099   & 0                  & -0.99       & 0.037      \\
del-sn099  & $\tau_{DF}$ (Eq. 
 \ref{dynamical_friction_time}) & -0.99       & 0.037      \\ \hline
\end{tabular}

\label{models}
\end{table}

\section{Calibrating against observations}
\label{sec:calibrate}
We calibrate the fiducial model against galaxy and AGN observables. The observables we consider are the UV luminosity (UVLF) of Lyman Break Galaxies (LBG) at redshifts $z \sim 5-14$ \citep{Bouwens2022Z2-9Turnover, Bouwens2021NewEfficiency, Harikane2022GOLDRUSH.10,Harikane2023AEpoch, Atek2015NewA2744, Atek2018TheUncertainties, Adams2022DiscoveryField, Adams2024EPOCHS.Data, Donnan2022TheImaging, Donnan2024JWST15, Bowler2017UnveilingTelescope/i, Leung2023NGDEEPImaging, Finkelstein2024The8.514.5, Willott2024AFunction, Robertson2024EarliestBang, Whitler2025TheReionization}, the stellar mass function (SMF) at $z \sim 5-10, 12$ \citep{Navarro-Carrera2024ConstraintsData, Song2016THEREDSHIFT, Duncan2014TheField, Bhatawdekar2019EvolutionFields, Stefanon2021GalaxyTime, Harvey2025EPOCHS.Observations}, the black hole mass function (BHMF) at $z \sim 5-6$ \citep{Kokorev2024AFields, Matthee2023LittleSurveys}, and the bolometric luminosity function at $z \sim 5-8$ \citep{Kokorev2024AFields, Akins2024COSMOS-Web:Assembly}. We deliberately exclude pre-JWST AGN data, as they predict $10-100\ $ times smaller number densities for the (extrapolated) bolometric LF at the faint end compared to JWST results. We also do not consider the ultraviolet LF, given the undetermined contribution from their galaxy host \citep{Perez-Gonzalez2024WhatEdition, Rinaldi2024NotDots}, or a possible contribution from scattered unreddened UV light from the AGN \citep{Greene2023UNCOVERz5, Kocevski2024TheFields, Kokorev2024AFields}. In a future paper, we will explore the observed UV LF compared to observations, assuming different attenuation laws and Hydrogen column densities. For the next subsections, we describe how we chose the best parameters for the fiducial model to reproduce JWST observables. Afterwards, we show how our models compare to the Bolometric LF and BHMF of AGNs. The errors for each model are calculated for each mass (luminosity) bin, assuming a Poisson Error weighted by the number density of the halos found at the corresponding bin, following the formalism presented in \citet{Bohm2013StatisticsApplications}. We refer the reader to Appendix \ref{best_fit_observables} for details in the calibrated galaxy observables.

\subsection{Finding the best parameters for the fiducial model}
\label{sec:finding_params}

To find the best parameters that simultaneously fit all observables, we perform a parameter exploration over all the free parameters in Delphi, namely: $ \{ f_{w}^{\star}$, $f_w^{\rm BH}$, $f_{\rm Edd} (M_h<M_h^{\rm crit})$, $f_{\rm Edd}(M_h \geq M_h^{\rm crit})$, $f_{\rm acc}^{\rm BH}(M_h<M_h^{\rm crit})$, $f_{\rm acc}^{\rm BH}(M_h \geq M_h^{\rm crit}) \}$. The range of values explored is detailed in Table \ref{table_param_Explore}, amounting to a total of approximately 6200 models. For each run, we construct the SMF and UVLF for LBGs and the BHMF and the bolometric LF for AGNs.

For each observable, we compute the chi-square statistic against available observations at multiple redshifts (see Table \ref{table:obser_chi2}). During this calculation, we avoid data points that are lower or upper limits\footnote{For most cases, our selected model agree with the lower and upper limits from observations.}. To characterize the overall performance for each observable, we sum the chi-square values across all redshifts, resulting in a chi-square distribution (see Fig. \ref{fig:chi2}) per observable (SMF, BHMF, UVLF of LBGs, and bolometric LF). 

To select the best-fitting parameters, we pre-select all models that fall within the 38th percentile of the chi-square distribution for each observable. We choose this specific percentile because it selects a small number ($N$=6) of models that show good overall performance across datasets. From this subset, we identify the model with the smallest total sum of chi-square values across all observables as our best-fit model. The best parameters are shown in Table \ref{parameters}.

\begin{table*}[]
\small
\caption{Best-fit free parameters for the fiducial model}
\centering
\begin{tabular}{lll}

\hline
Parameter                         & Best fitting values & Description                                           \\ \hline
$f_{\mathrm{acc}}^{\mathrm{bh}}$\tablefootmark{a} & 0.1 (5$\times$10$^{-3}$)\tablefootmark{b}  & Fraction of gas accreted by the BH after SNII feedback \\
$f_{\mathrm{Edd}}$\tablefootmark{a}               & 1.0  ($10^{-3}$)\tablefootmark{b}      & Eddington fraction accreted by the BH.                \\
$f_w^{\mathrm{bh}}$               & 5$\times$10$^{-4}$           & Fraction of BH feedback coupled to winds.             \\
$f_w^{\star}$                     & 0.04                 & Fraction of SNII feedback coupled to winds   \\ \hline   
\end{tabular}
\tablefoot{A detailed explanation on how we found these values can be found in Appendix \ref{Appendix:best_fit_params}.
\tablefoottext{a}{These parameters are implemented with a 0.5 dex scatter} \tablefoottext{b}{Values in parenthesis correspond to the case when the halo lies below the critical threshold $M_h^{\mathrm{crit}} = 10^{11.25} [\Omega_m (1+z)^3 + \Omega_{\Lambda}]^{-0.125} M_{\odot}$, which also has a 0.5 dex scatter.}}
\label{parameters}
\end{table*}

\subsection{Bolometric Luminosity Function}

\label{subsec:BolLF}

\begin{figure*}
    \centering
    \includegraphics[width=500pt]{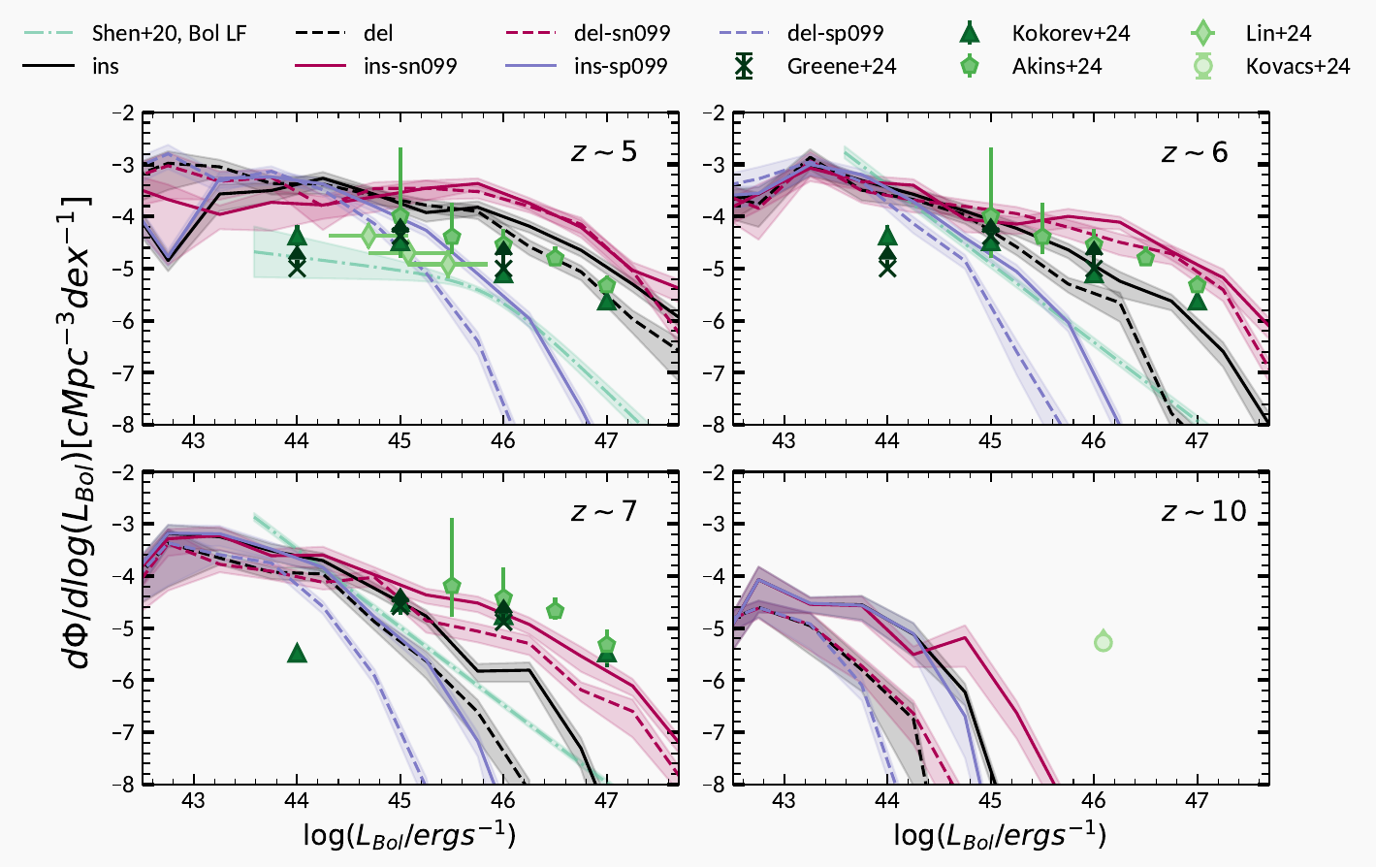}
        \caption{Bolometric luminosity functions for AGNs for redshifts 5, 6, 7 and 10. The instantaneous models are shown as solid lines, while the delayed are dashed. The shaded areas for each model correspond to their Poisson errors. The fiducial model (ins) and delayed ({\sc del}) are shown in black, the extreme retrograde are shown in pink and the extreme prograde in blue (del-sp099). Observations pre-JWST are represented by the fits from \citet{Shen2020The07} (green dash-dot lines). JWST results are from \citet{Greene2023UNCOVERz5,Kokorev2024AFields, Matthee2023LittleSurveys, Kovacs2024Azapprox10, Akins2024COSMOS-Web:Assembly, Lin2024AWFSS}.}
    \label{fig:lbol}
\end{figure*}

The Bolometric luminosity function (bolometric LF) is shown in Fig. \ref{fig:lbol} for $z \sim 5-7$ and $10$. We remind the reader that only the fiducial model (black solid lines) have been calibrated against observational data. The shaded regions around each theoretical model show their Poisson errors, weighted by the number density of BHs with a bolometric luminosity corresponding to each bin. The del-sp099 and ins-sn099 models correspond to the lower and upper bounds of BH growth and luminosity, respectively. Observations pre-JWST are represented with the LF fits from \citet{Shen2020The07} , shown as the gray dot-dashed lines. JWST observations at redshifts $5-10$ come from \citet{Greene2023UNCOVERz5}, \citet{Kokorev2024AFields}, \citet{Matthee2023LittleSurveys}, \citet{Akins2024COSMOS-Web:Assembly}, \citet{Lin2024AWFSS} and \citet{Kovacs2024A10}. 

We summarize how each observational work estimated AGN bolometric luminosities, highlighting their key assumptions and caveats. \citet{Greene2023UNCOVERz5} selected high-$z$ (z>4) AGN candidates via photometry based on the little red dot (LRD) criteria, being compactness and a characteristic "v"-shaped SED. They confirmed the presence of an AGN by detecting broad H$\alpha$ in their NIRSpec spectra. Assuming the observed UV corresponds to scattered (or leaked) AGN emission, they correct for dust using the observed UV slope and estimate the bolometric luminosities using standard relations \citep{Greene2005EstimatingLine}.  \citet{Kokorev2024AFields} also used the LRD photometric criteria and performed a two-component SED fitting, where the UV/Optical comes from the Sloan Digital Sky Survey (SDSS) composite quasar SED \citet{VandenBerk2001CompositeSurvey} and the NIR comes from \citet{Glikman2006AQuasars}. In this template, they consider a dust reddened SED alongide a non-attenuated emission arising from scattered light. The bolometric luminosity is derived using the luminosity at $5100$\r{A} and the corresponding standard relations \citep{Greene2005EstimatingLine, Kaspi2000ReverberationNuclei}. 
\citet{Akins2024COSMOS-Web:Assembly} photometrically selected compact red sources and fitted quasar-only SEDs with up to 10$\%$ scattered, unattenuated UV emission. Bolometric luminosities were derived from the monochromatic luminosity at 3000\r{A} of their best SED fit, using the bolometric corrections from \citet{Richards2006SpectralQuasars}.
\citet{Lin2024AWFSS} photometrically searched for broad H$\alpha$ in compact, red sources, from which they estimated $L_{5100}$ \citep{Greene2005EstimatingLine} and the corresponding bolometric luminosity \citep{Richards2006SpectralQuasars}. They do not apply dust corrections due to their limited spectral coverage, potentially underestimating the bolometric LF by 0.4-0.8 dex if $A_V = 2-4$. Finally, \citet{Kovacs2024A10} estimated the bolometric LF using two of $z \sim 10$ galaxies \citep{Kovacs2024Azapprox10, Bogdan2023EvidenceQuasar}, identified via CHANDRA soft X-ray emission. The bolometric luminosities were derived using \citet{Lusso2012BolometricSurvey} corrections.

To reproduce the JWST-AGN bolometric LF while maintaining consistency with galaxy observables, our models require accretion parameters of $f_{\rm Edd} = 1\ (10^{-3})$ and $f_{\rm acc}^{\rm BH}=0.1\ (5 \times10^{-3})$ in halos above (below) the critical threshold. It is important to consider that in D19, the accretion parameter $f_{\rm acc}^{\rm BH}$ did not have two different values depending on the critical mass threshold, as BH growth in low-mass halos was effectively suppressed. Hence, our current prescription extends the framework by allowing low mass halos to contribute on BH growth via accretion whenever there is sufficient gas. 
Compared to D19, which was calibrated to pre-JWST high-$z$ AGN surveys, our accretion parameters above the critical halo mass threshold are $\approx$ 1$-$2 orders of magnitude higher \footnote{If we were to use the former theoretical framework of {\sc Delphi} we find that the accretion parameters must be $\approx$ 2$-$3 orders of magnitude higher}. This means that BHs in large halos ($M_h\geq M_h^{\rm crit}$) must at least accrete $10\%$ of the remaining gas after SNII feedback, i.e. $\sim 100$ times more gas than in D19. At the same time, in our updated framework, BHs in small halos ($M_h<M_h^{\rm crit}$) have to accrete at least $\sim 0.1\%$ of gas after stellar processes.
 Both D19 and this work impose Eddington limited ($f_{\rm Edd}=1$) growth for high mass halos ($M_h>M_h^{\rm crit}$). It is important to note that in D19, the parameters for BH accretion and the critical halo mass threshold ($M_h^{\rm crit}$) for a given halo were not implemented with a 0.5 dex gaussian scatter in log-scale, hence we achieve a smoother transition between fractions that yield high ($f_{\rm Edd}=1$ and $f_{\rm acc}^{\rm BH}=0.1$) and low accretion ($f_{\rm Edd}=10^{-3}$ and $f_{\rm acc}^{\rm BH}=5 \times 10^{-3}$).

Nonetheless, incrementing the parameters for BH accretion, as well as setting heavy initial BH masses results in strong BH feedback, which is counteracted by a weaker value of BH feedback coupling onto winds ($f_w^{BH}=5 \times 10^{-4}$), of about an order of magnitude lower than in D19, reporting $f_w^{BH}= 0.003$. Consequently, this elevated accretion accelerates the assembly of heavy, $10^6-10^9\ M_{\odot}$ accreting BHs, shining with luminosities between $10^{42}- 10^{47}$ erg s$^{-1}$, which boosts the number densities of BHs in this mass and luminosity regime. Additionally, the incorporation of {\sc Sphinx}$^{20}$ cold gas fractions and star formation efficiencies results in a drastic diminution of SNII feedback limited halos (i.e. halos with no gas after SN feedback). Before incorporating the values from {\sc Sphinx}$^{20}$, in {\sc Delphi}, $96.06\%$ ($99.92\%$) of halos at redshift 5 (10) were completely devoid of gas following stellar and BH feedback. After the update, this fraction has significantly decreased to $0.11 \%$ ($9.63\%$) for redshifts 5 (10). This change primarily facilitates the growth of black holes in low-mass halos by allowing continued gas accretion, thereby accelerating their assembly.

We now discuss the evolution of the bolometric LF shown in Fig. \ref{fig:lbol}, moving from high to lower redshifts. Across all panels, the interplay between BH accretion rates, spin configurations, and merging prescriptions determines the shape and normalization of the LF, though their relative importance evolves with redshift.

At $z$=10 (bottom-right panel), models with instantaneous mergers (ins, ins-sn099, ins-sp099) reach higher number densities across the entire luminosity range compared to the delayed models (del, del-sn099, del-sp099). This stems from the accelerated BH growth by immediate BH-BH coalescence; this allows them to grow even faster once their halos are massive enough to retain gas after SNII feedback. Most BHs in this luminosity regime ($\geq 10^{42.5}$ erg s$^{-1}$) accrete in the $f_{\rm Edd}$-regulated mode, where accretion scales with BH mass. Additionally, the impact of spin is evident: more negative spins result in higher number densities at the bright end $\geq 10^{44.4}$erg s$^{-1}$ ($10^{43.2}$ erg s$^{-1}$) of the bolometric LF for instantaneous (delayed) models. The latter is due to the low radiative efficiency and thus, higher mass accretion rates that more negative spins yield (see Eq. 2). However, none of our models approach the high number density estimated by  \citet{Kovacs2024Azapprox10}, with even our upper bound model (ins-sn099) underpredicting this prediction by more than 5 dex. We interpret this rare, bright sources as outliers relative to the AGN population our models aim to describe.  
At $z$=7 (bottom-left), spin strongly modulates the shape of the bolometric LF. Models with extreme retrograde spin (ins-sn099, del-sn099) yield the highest number densities at luminosities $>10^{45}$erg s$^{-1}$, followed by the canonical and extreme retrograde spin. This hierarchy reflects how more negative spins boosts accretion rates and hence BH mass and luminosity. Instantaneous models remain consistently more number dense than their delayed counterparts at fixed spin due to their enhanced merger-driven growth and gas availability (supplemented from satellite galaxies). Interestingly, at fainter luminosities $<10^{43.8}$ erg s$^{-1}$, extreme prograde spin models (ins-sp099 and del-sp099) report slightly higher number densities, as their slow growth via accretion leads to an abundant population of low-luminosity BHs. When compared to JWST data \citep{Akins2024COSMOS-Web:Assembly, Kokorev2024AFields, Greene2023UNCOVERz5}, models with extreme retrograde spin (ins-sn099 and del-sn099) match the bright-end number densities $\approx 10^{-5} - 10^{-6}$ cMpc$^{-3}$ dex$^{-1}$ at $10^{45}- 10^{46}$ erg s$^{-1}$. The fiducial model agrees well up to luminosities of $10^{45}$erg s$^{-1}$, but under-predicts brighter bins by 0.9 to 3.2 dex. Miss-matches on some datasets are expected since the best-fit parameters are tuned to match both galaxy and AGN observables simultanously. 
At $z$=6 (top-right), both spin and merging prescriptions, particularly at luminosities $\gtrsim  10^{44}$ erg s$^{-1}$, determine the shape of the LF. The order in which models reach higher number densities for brighter luminosities remains consistent with that of $z$=7, highlighting the sustained role of spin-enhanced accretion.
 At the faint end, however, delayed and instantaneous models overlap within uncertainties. This convergence is driven by the following: delayed models maintain high accretion rates ($f_{\rm Edd}^{\rm eff}>0.5$) for small BHs ($\approx 10^{5-6.5} M_{\odot}$) due to their weaker feedback retaining more gas. Whereas, instantaneous models rapidly assemble massive BHs $\approx 10^{7.5-8} M_{\odot}$, which yield stronger feedback that effectively restricts their gas supply, lowering accretion rates ($f_{\rm Edd}^{\rm eff} < 0.1$) and thus populating the faint end of the bolometric LF. Delayed models also benefit from increased merger rates at this redshift, enabling BH growth via mergers as well as further gas accretion from the gas provided by the incoming satellite galaxy. The fiducial model shows its best match to observations at $10^{45 - 46}$ erg s$^{-1}$, slightly under-predicting the data by up to 0.7 dex at brighter luminosities.

By $z$=5 (top-left), the LF shows a narrowing gap between delayed and instantaneous models at fixed spin, particularly in the extreme retrograde case (ins-sn099 vs del-sn099), which are nearly indistinguishable across $10^{44.3-47.2}$ erg s$^{-1}$. This convergence arises from the increasing merger rates that delayed models experience at lower redshifts, which grow their BHs via BH-BH mergers as well as replenishes the available gas for further accretion. Consequently, delayed models "catch up" to instantaneous ones in bolometric luminosities, especially for spin configurations that maximize accretion. Interestingly, while most models with the same merging prescription overlap at the faint end ($\leq 10^{44}$ erg s$^{-1}$), ins-sn099 displays a distinct convex shape due to the buildup of $\sim 10^9 M_{\odot}$ BHs from prolonged, high accretion. Comparing to observations, the fiducial model now overpredicts the data by 0.22-1.6 dex depending on the luminosity bin. This suggests that models matching well at $z$=6-7 may overestimate the AGN population at later-epochs, possibly reflecting on observational incompleteness of $z$=5 AGNs, redshift evolution of AGN obscuration, or model limitations of BH quenching at later times ($z$<6).

\subsection{BH mass function}
\label{subsec:bhmf}

\begin{figure}
    \centering
    \includegraphics[width=\columnwidth]{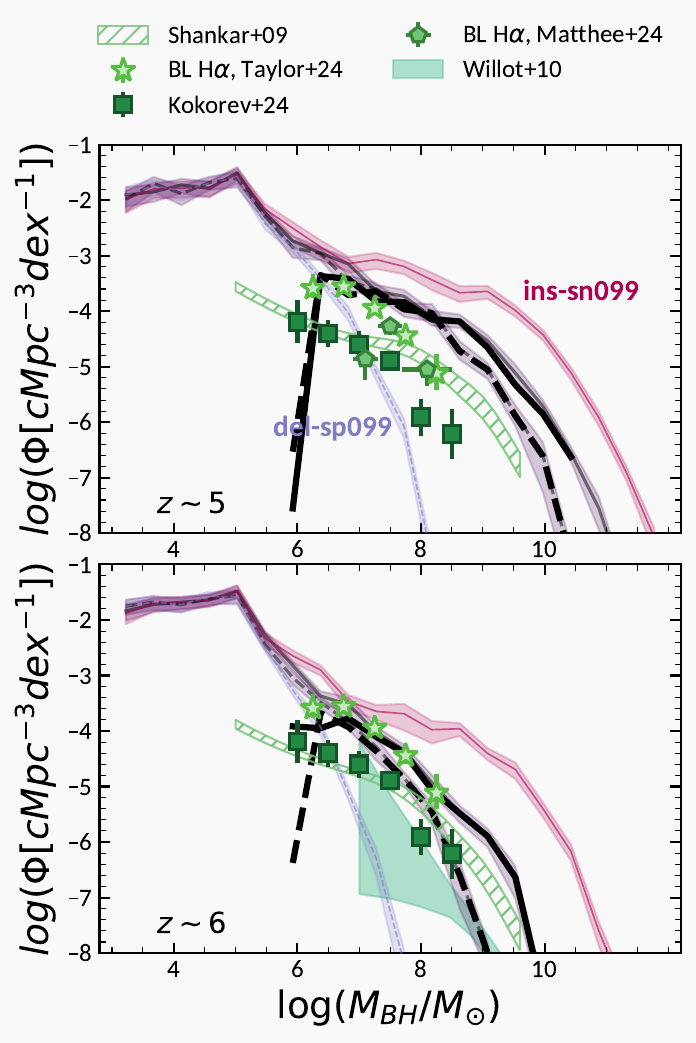}
    \caption{Black Hole mass functions at redshifts 5 and 6 for the fiducial (black solid lines), {\sc del} (black dashed lines), ins-sn099 (pink solid lines), del-sp099 (blue dashed lines). The thin lines represent all BHs, while thick lines show BHs with $L_{Bol} \geq 10^{44}$ erg s$^{-1}$ and $f_{\rm Edd}^{\rm eff}>0.1$. Observational constraints from Pre-JWST come from \citet{Willott2010EDDINGTON-LIMITED6} and \citet{Shankar2009Self-consistentEfficiency}, while JWST predictions are taken from \citet{Kokorev2024AFields}, and \citet{Matthee2023LittleSurveys}.}
    \label{bhmf}
\end{figure}

In Fig. \ref{bhmf}, we show the BHMF for redshifts 5 and 6. The ins-sn099 and del-sp099 models correspond to the upper and lower bounds of BH growth. The faint and thin lines correspond to the mass function of all BHs, while for the fiducial and {\sc del} model we show the BHMF for BHs $\geq 10^{44}$ erg s$^{-1}$ with $f_{\rm Edd}^{\rm eff}\geq 10\% $ in the thick and saturated lines. The choice for the latter constraint is motivated by the faintest luminosity and Eddington fractions probed from spectroscopic JWST observations \citep{Matthee2023LittleSurveys,Greene2023UNCOVERz5}. 

Pre-JWST predictions for the BHMF come from \citet{Shankar2009Self-consistentEfficiency}, while direct BHMF estimations are from \citet{Willott2010EDDINGTON-LIMITED6}. 
The former derives the BHMF using a reference model of BH evolution where BHs accrete at 0.6 Eddington fraction and a radiative efficiency of $\epsilon$=0.065. They calibrate the BHMF so it reproduces the Bolometric Luminosity functions of AGNs available at that time. In the panel at redshift 5 we show \citet{Shankar2009Self-consistentEfficiency} reference model for redshifts 4.5-5, while at redshift 6 we show their predictions between redshifts 5.51-5.99. Meanwhile, the latter estimates the BHMF so it can reproduce the luminosity functions in \citet{Willott2010The6}, considering an intrinsic Eddington distribution for $z\approx 6$ sources, an obscured fraction and a duty cycle of $\sim 0.75$. Regarding JWST observations, \citet{Matthee2023LittleSurveys} constructed the BHMF for redshift bins $4-6$ using single-epoch measurements of broad H$\alpha$ in their LRD sample. \citet{Kokorev2024AFields} estimated the BHMF at redshift bins $4.5-6.5$ for their photometrically selected LRD sample using their bolometric LF and assuming that all their sources are shining at the Eddington limit. Finally, \citet{Taylor2025Broad-LineDots} constructed the BHMF at redshift bins $3.5-6$ by calculating the single-epoch BH mass from their sources with detectable broad H$\alpha$ in slitless spectroscopy. Out of this sample, they find that $\sim$30$\%$ satisfy the LRD criteria. Given the complexity of SED modeling, they decide not to incorporate corrections from dust attenuation.

At redshifts 5 and 6, we see a somewhat flat distribution of BHs between $10^3 - 10^5 M_{\odot}$ masses for all displayed models. This shape comes from the log-uniform distribution of BH seeds seeded at starting halos down to redshift $\approx 13$. Through cosmic time, these BHs are starving or accreting at sub-Eddington fractions in small halos ($M_h < M_h^{crit}$), which easily eject all their gas from stellar and BH feedback. In the high-mass counterpart, the overall shape of the BHMF is mainly subject to spin prescription, i.e., the accretion rate and the gas availability.  
Meanwhile, in the case of models with canonical spin, we see that the delayed and instantaneous models overlap within error bars for BH masses below $10^{8.5} M_{\odot}$ ($10^{7} M_{\odot}$) for redshifts 5 (6). Indeed, through instantaneous merging, the number density of massive BHs ($\geq 10^8 M_{\odot}$) are at least 0.3 dex more numerous than the delayed, due to efficient merging and accretion from the maximum amount of gas possible brought in by satellite galaxies.

Comparing to JWST observations, we see that at redshift 6 (bottom panel of Fig. \ref{bhmf}), the fiducial model finds good agreement with the data points from \citet{Taylor2025Broad-LineDots}, while for \citet{Kokorev2024AFields} we overpredict the number densities by 0.7$-$1.27 dex. Even after masking by Bolometric Luminosity ($L_{\rm Bol}\geq10^{44}$erg s$^{-1}$) and Eddington fraction ($f_{\rm Edd}\geq10\%$), we overpredict \citet{Kokorev2024AFields} points by 0.26$-$1.11 dex. Whereas at redshift 5, the masked BHMF of the fiducial agrees with \citet{Taylor2025Broad-LineDots} between $10^{6.25} - 10^{7.25} M_{\odot}$, and overpredicts by 0.59$-$1 dex at higher masses, while for \citet{Matthee2023LittleSurveys} and \citet{Kokorev2024AFields} we overpredict the number densities at all mass ranges by 0.43$-$1.19 dex and 1$-$1.9 dex, respectively. Regarding pre-JWST observations, as expected, we overpredict the number density by at least 1 dex at all mass ranges for redshifts 5 and 6.

Although the fiducial model tends to overestimate most data points, it is important to consider that the BHMF estimation from \citet{Kokorev2024AFields} does not account for the underlying distribution of Eddington fraction, which is set to 1. Therefore, assuming an average Eddington fraction of $10 \%$ would shift these data points to the right by 1 dex, comfortably matching the BHMF of the fiducial model. On the other hand, both BHMF estimated from \citet{Matthee2023LittleSurveys} and \citet{Taylor2025Broad-LineDots} are built using single-epoch measurements of broad H$\alpha$, which are uncertain to be valid for sources accreting above or close to the Eddington limit \citep{Lupi2024SizeJWST,King2024TheQSOs}. Additionally, they do not correct for dust attenuation, hence, the estimations for BH masses may be underestimated. At the same time, accounting for type II AGN may increase the number density by up to 0.6 dex \citep{Maiolino1995Low-LuminosityGalaxies}, if the proportion of type II over type I found at low-z stands at these high-$z$.

\section{Emergence of Black Holes in the first Gyr}
\label{sec:emergence_bhs}

In this section, we will describe and discuss the emergence of BHs and their host galaxies from 0.125 Gyr ($z=26$) up to 1.2 Gyr of the Universe ($z=5$), spanning the first billion years in cosmic time. We will compare the configurations presented in this work (see table \ref{models}) and analyze their impact on the stellar to BH mass relation, as well as their expected merger rates. 

\subsection{Stellar and BH mass relation}

\begin{figure*}[ht]
    \centering     
    \includegraphics[width=510pt]{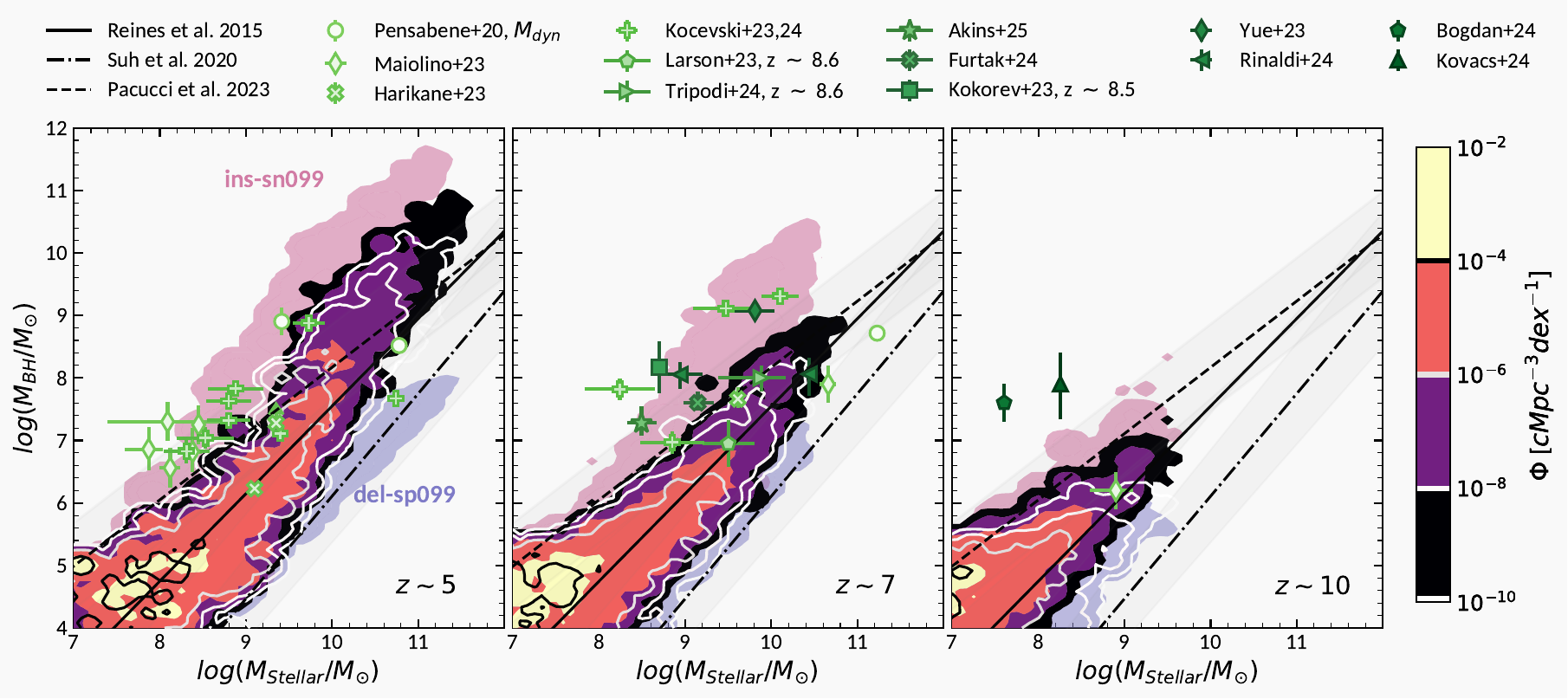}
    \caption{ Stellar to black hole (BH) mass relation for the fiducial (shaded contours) {\sc del} (line contours), ins-sn099 and del-sp099 models at redshifts 5, 7 and 10. Here, we present the relation as contours from a smoothed 2d histogram, weighted by their associated number density. In all panels, we show the scaling relations for local Elliptical and Spheroidals (solid black line) \citep{Reines2015RelationsUniverse}, high redshift galaxies ($z \sim 2.5$) \citep{Suh20202.5} (black dash-dotted line), and the relation for high-redshift (z=4-7) AGNs using only JWST sources  \citep{Pacucci2023JWSTModels} (black dashed lines). Observations pre-JWST come from works of \citet{Pensabene2020TheGalaxies}, and the latest results from JWST from \citet{Maiolino2023JADES.Mighty}, \citet{Harikane2023JWST/NIRSpecProperties}, \citet{Yue2023EIGERzgtrsim6}, \citet{Bogdan2023EvidenceQuasar}, \citet{Rinaldi2024NotDots}, \citet{Tripodi2024RedMyr}, \citet{Larson2023AQuasars}, \citet{Kokorev2023UNCOVER:8.50}, \citet{Akins2025StrongFormation} and \citet{Kovacs2024A10}.}
    \label{smvsbhm}
    
\end{figure*}

From observations we have BH and stellar masses that range between $\log(M_{\rm BH}/M_{\odot}) = [6-9.5]$ and $\log(M_{\rm stellar}/M_{\odot}) = [7.5-11.6]$, and from redshifts 5 to 10. Most BH masses were estimated using single-epoch measurements of broad H$\alpha$ \citep{Harikane2023JWST/NIRSpecProperties, Maiolino2023JADES.Mighty,  Kocevski2023HiddenCEERS, Kocevski2024TheFields, Akins2025StrongFormation, Rinaldi2024NotDots}, H$\beta$ \citep{Yue2023EIGERzgtrsim6, Kokorev2023UNCOVER:8.50, Furtak2023AShadows, Tripodi2024RedMyr, Kocevski2024TheFields}, or MgII \citep{Pensabene2020TheGalaxies} depending on the spectral coverage. Other works \citep{Kovacs2024Azapprox10,Bogdan2023EvidenceQuasar} derive BH masses from X-ray emission found in {\sc chandra}, assuming an Eddington limited accretion and X-ray bolometric corrections \citep{Lusso2012BolometricSurvey, Duras2020UniversalDecades}.
 The stellar masses for some of these observations are estimated by SED fitting the photometric points of host-only images \citep{Harikane2023JWST/NIRSpecProperties, Ding2023Detection6, Yue2023EIGERzgtrsim6}, i.e. on the residual images after subtracting the PSF contribution of the quasar. Other stellar masses are derived from SED fitting the photometric and/or spectroscopic data of the complete source with \citep{Maiolino2023JADES.Mighty, Kokorev2023UNCOVER:8.50, Kocevski2023HiddenCEERS, Kocevski2024TheFields, Tripodi2024RedMyr, Akins2025StrongFormation, Rinaldi2024NotDots} and without \citep{Bogdan2023EvidenceQuasar, Kovacs2024Azapprox10, Furtak2023AShadows} an AGN component. For the case of \citet{Pensabene2020TheGalaxies}, they measure the dynamical mass using the CII interstellar medium tracer from ALMA observations.

Starting at redshift 10, in the right panel of Fig. \ref{smvsbhm}, we see galaxies and BHs in the fiducial model scattered along a slightly shallower slope compared to the local relation for Elliptical and Spheroidals \citep{Reines2015RelationsUniverse}. Around $8.35 \%$ of galaxies $M_{\rm stellar}  \lesssim 10^{8} M_{\odot}$ hold non-accreting BHs. These systems are located in small halos that easily expel all their gas via SNII feedback. The rest of these galaxies at this regime and higher have sufficient gas (after stellar feedback) for their BH to accrete, at a fraction that increases towards higher stellar masses. The largest BHs ($\approx 10^7 M_{\odot}$) at this time are located in the rarest and most massive halos ($10^{-10}$ cMpc$^{-3}$). For {\sc del} we find a similar trend but with a shallower slope than the fiducial, scattered down towards smaller BH masses. This results from BHs in the {\sc del} model having a slower growth via mergers, consequently resulting in smaller initial masses once they start accreting. Unlike the fiducial model, the rarest halos in {\sc del} achieve BH masses that barely exceed $10^6 M_{\odot}$. We also note how the addition of a dynamical timescale for BH and galaxy mergers becomes more impactful for growing massive BHs ($>10^6 M_{\odot}$) rather than massive galaxies. This can be understood since star formation processes occur in halos at all redshifts and mass ranges, as opposed to BHs only being present in starting halos down to $z=13$, with efficient growth via accretion only at high mass systems ($M_h \geq M_h^{\rm crit}$). 

The observational data points available at this extreme redshift are from \citet{Maiolino2023JADES.Mighty}, and gravitationally lensed galaxies \citep{Bogdan2023EvidenceQuasar, Kovacs2024Azapprox10}. The latter two works report overmassive BHs with respect to their host galaxy, comprising $\approx 10 \%$ of the stellar component, as opposed to $\approx 0.01\%$ from local relation predictions \citep{Reines2015RelationsUniverse}. We can only obtain systems such as the one from \citet{Maiolino2023JADES.Mighty}, while we fail to reproduce sources as in \citet{Bogdan2023EvidenceQuasar} and \citet{Kovacs2024Azapprox10}. Not even our model with extreme retrograde (ins-sn099) can reach such extreme sources. As mentioned in the previous sections, our theoretical framework aims to reflect upon the general population of BHs rather than these extreme cases. Nonetheless, a potential mechanism that can help to reach these elevated masses is to include prescriptions of Super-Eddington accretion \citep{Lupi2024SizeJWST}. 

Around 300 Myrs later, at $z\sim 7$, as shown in the middle panel of Fig. \ref{smvsbhm}, the fiducial model is scattered at a comparable slope with local relations for Ellipticals and Spheroidals, reaching BH masses of $\geq 10^9 M_{\odot}$ in massive galaxies ($\geq 10^{11}M_{\odot}$), located at the rarest halos ($\leq 10^{-10}$ cMpc$^{-3}$). At this time, $1.14 \%$ of galaxies below $10^{8.29} M_{\odot}$ harbor non-accreting BHs driven by growth via mergers, while the rest additionally grow from accretion. Meanwhile, the {\sc del} model is scattered at a slightly shallower slope relative to the fiducial, growing BHs up to $\sim 10^{8.5} M_{\odot}$ in the rarest halos ($\leq 10^{-10}$ cMpc$^{-3}$). The steady increase of merging halos in the {\sc del} model, allows the main system to increase its stellar, and BH components (if a BH is present) from satellite galaxies. Furthermore, as shown in previous work \citep{Piana2020TheHoles, Piana2021TheGalaxies}, BH accretion becomes the main driver of BH growth in massive haloes that surpass the critical halo mass threshold. At this stage, accretion contributes up to an order of magnitude more mass than mergers. The extent to which accretion dominates over mergers depends on the halo's individual merger history and its final mass at the end of the simulation. For instance, in the fiducial, haloes that grow to $\rm log (M_h/ M_{\odot})=10.5-11$ by z=4.5 host BHs that stochastically grow from mergers, with accretion contributing an average of 37$\%$ of the total BH mass by z$\sim$5. In contrast, halos that reach $\rm log (M_h/ M_{\odot})=12.5-13\ (13.5-13.9)$ by z=4.5 host BHs whose masses are on average 98 (91)$\%$ of their BH mass built through accretion events.
We refer the reader to \citealt{Piana2020TheHoles, Piana2021TheGalaxies} for a detailed analysis of the relative contributions of mergers and accretion across redshifts. Notably, in the current model, where SN feedback-limited systems are far less common than in earlier versions of the {\sc delphi}, most satellite galaxies retain sufficient gas to further BH accretion.

Regarding observations, we find the fiducial model in agreement with only some JWST data points \citep{Yue2023EIGERzgtrsim6,Maiolino2023JADES.Mighty, Harikane2023JWST/NIRSpecProperties, Kocevski2024TheFields, Larson2023AQuasars, Tripodi2024RedMyr}, while for the rest we are unable to reproduce such overmassive BHs at a given stellar mass \citep{Kokorev2024AFields, Kocevski2024TheFields,  Akins2025StrongFormation}. However, we do note that many of the data points that the fiducial does not reproduce are within the maximum limit of BH growth from the extreme retrograde model (ins-sn099).

Roughly 500 Myrs later, at redshift 5 (left panel of Fig. \ref{smvsbhm}), we note that BHs in galaxies above $10^{9.8} M_{\odot}$ in the fiducial model experienced significant growth with respect to the previous redshift. At this point, $2.24 \%$ of galaxies $\lesssim 10^{10.81} M_{\odot}$ have no gas after BH and stellar feedback. Meanwhile, most BHs in galaxies with stellar mass above $10^9 M_{\odot}$ lie below the 0.69 dex scatter of the JWST high-$z$ relation \citep{Pacucci2023JWSTModels}, but $7.05 \times 10^{-5}$ cMpc$^{-3}$ ($0.16\%$) of them are found above the 0.55 dex scatter of the local relation from Ellipticals and Spheroidals \citep{Reines2015RelationsUniverse}. Although these overmassive BHs ($10^8 - 10^{10} M_{\odot}$) are a minority ($<1\%$ of the population), the sustained elevated accretion rates pose a complicated scenario for these systems to agree with local ($z\approx0$) relations. Under our cosmological constants, at redshift 5 the Universe has a comoving volume of 2111.765 Gpc, which means that our most extreme BH sources $\geq 10^{10}M_{\odot}$ associated to the rarest number densities $ \lesssim 10^{-6}$ cMpc$^{-3}$ lead to a maximum number of $\approx 2 \times 10^7$ overmassive sources. Therefore, additional mechanisms are required to quench effectively the BH growth in high mass systems ($M_{\rm stellar}>10^{10} M_{\odot}$), than those already incorporated in the $f_w^{\rm BH}$ parameter that regulates the wind coupled energy. 

Concerning the differences between the fiducial and {\sc del} model demographics, these are indistinguishable for stellar masses below $10^{10} M_{\odot}$ harbouring BHs $\geq 10^8 M_{\odot}$ (also reflected in Fig. \ref{bhmf}). The increase of mergers in the {\sc del} model allows small galaxies to grow their BHs from mergers, whereas satellites merging in large galaxies would additionally provide a reservoir of gas that boosts star formation and BH accretion.
Again, the fiducial agrees with some observations except those found in \citet{Maiolino2023JADES.Mighty, Kocevski2024TheFields}, which notably deviates from local relations, while partially cross-matching within the 0.69 dex scatter from the proposed relation from \citet{Pacucci2023JWSTModels} for high-$z$ JWST-AGNs. Nonetheless, our model of extreme BH growth (ins-sn099) is able to cover the overmassive BHs from \citet{Kocevski2024TheFields}, while only partially reaching those from \citet{Maiolino2023JADES.Mighty}.

\subsection{Merger rates in the first Gyr}
\label{sec:merger_rates}

In this section, we calculate the raw merger rates for BHs in all the models explored in this work (see Table \ref{models}), i.e. without filtering by the sensitivity of future gravitational observatories like LISA or ET. We compare our results with previous predictions found in literature. We begin calculating the merger rates as a function of redshift, using the following expression \citep{Arun2009MassiveTaskforce}:

\begin{align}
    \frac{d^2 N}{dz dt } = \frac{4 \pi c}{\Delta z} \sum_{N \in \Delta z} n(z) \left( \frac{d_L(z)}{1+z} \right)^2 [yr^{-1}]
    \label{Eq:merger_fnz_z}
\end{align}

\noindent where $n(z)$ is the number density of mergers at redshift $z$ (derived from the host halo), and $d_L$ is the luminosity distance at a given redshift. Since the merging timescale for galaxies and BHs solely depends on the mass ratios between their dark matter halos, varying the values for BH spin has no impact on the merger rates as a function of redshift. Therefore we show the merger rates as a function of $z$ distinguished by the merging timescale only.

\begin{figure}
    \centering
    \includegraphics[width=\columnwidth]{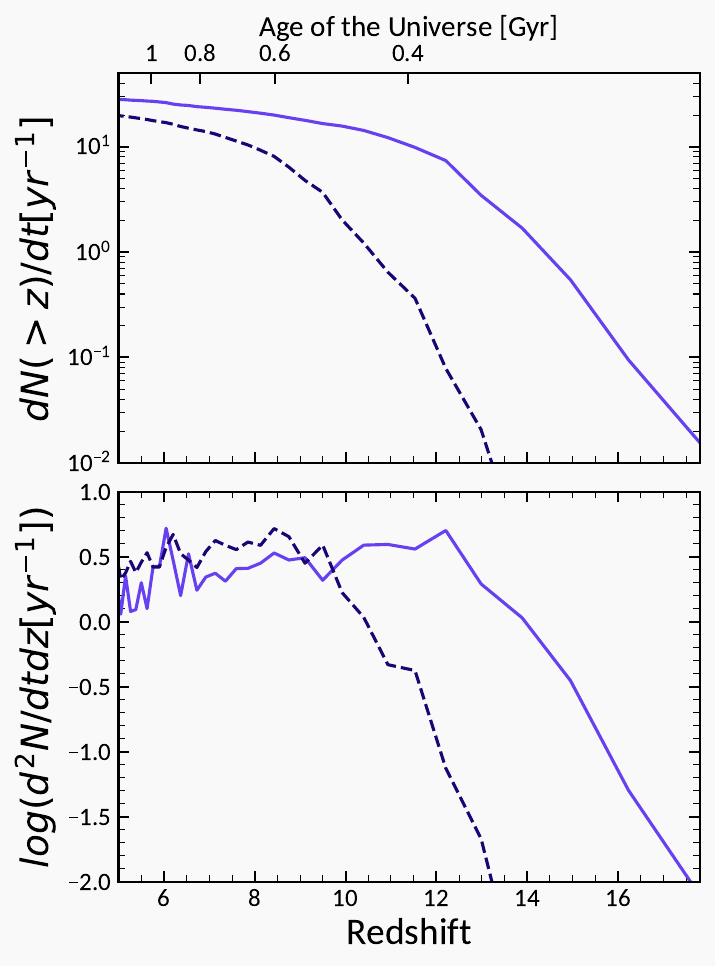}
    \caption{In the upper panel, we show the cumulative merger events for BHs per year as a function of redshift, and in the bottom are the merger rates at a given redshift bin. Instantaneous models are shown as solid lines, while delayed are dashed.}
    \label{fig:merger_rates_bhs}
\end{figure}

In Fig. \ref{fig:merger_rates_bhs} we show in the upper panel, the cumulative raw merger rates in units of yr$^{-1}$ up to redshift 5 for all BHs in the instantaneous (solid line) and delayed models (dashed line), while in the lower panel we show the merger rates for a given redshift bin as a function of redshift (Eq. \ref{Eq:merger_fnz_z}). In the upper panel we see that the instantaneous models reports a total of 28.06 merger events per year for $z \geq 5$, while the delayed reports 19.61 mergers per year. In the lower panel, we note that the delayed models start to merge more frequently ($> 1$ yr$^{-1}$) at redshift below 10.5, whereas the instantaneous report such frequencies as early as redshift 14.

Merger rate predictions for heavy BH seeds are explored in previous works in the literature \citep{Sesana2007TheStream, Bonetti2019Post-NewtonianLISA, Barausse2020MassiveFeedback, Klein2016ScienceBinaries, Dayal2019TheAstronomy}. Most of these estimations are based on the formation of heavy seeds from dynamical instabilities as presented in \citet{Begelman2006FormationHaloes} and \citet{Volonteri2008TheSeeds}, where a heavy seed is formed after the rapid growth of a quasi-star, or as in \citet{Koushiappas2004MassiveMaterial}, where large BHs form as a Direct Collapse within halos with gas regions of low angular momentum. \citet{Dayal2019TheAstronomy}, on the other hand, explores the formation of DCBHs in Atomic Cooling Halos (ACHs) exposed to a critical Lyman Werner (LW, at $11.2-13.6$ eV) background. We caution the reader that our seeding mechanism for BHs only requires metal free starting halos found at reshifts $\approx$ 13 and above, hence the number density of seeds is subject to the resolved $10^8 M_{\odot}$ halos in pur DM merger tree halo. For a more detailed comparison between the number densities of the BH seeds in other works, we refer to the Appendix \ref{Appendix:number_densities}.

From \citet{Sesana2007TheStream}, we find merger rates ranging from 2 to 182 yr$^{-1}$ at $z > 5$ when varying the efficiency of metal enrichment and the prescriptions for heavy seed formation. Similarly, we find that \citet{Klein2016ScienceBinaries} obtains $3 - 88$ yr$^{-1}$ events at $z > 5$ considering delayed and instantaneous merging. \citet{Bonetti2019Post-NewtonianLISA} finds agreement with the works mentioned, with a total of $\sim 23$ yr$^{-1}$ mergers for heavy seeds at $z > 0$ in their delayed model, including sub-kpc evolution of binary BHs within a stellar or gaseous environment, including the interaction with a third possible BH. From \citet{Barausse2020MassiveFeedback}, we find merger rates ranging from 0.2 - 2441 yr$^{-1}$ at $z > 5$ in their models with SN feedback, while exploring different delay timescales and instantaneous merging. \citet{Katz2020ProbingLISA} report a range of $0.007-75$ yr$^{-1}$ events for BH seeds in the $10^{5}-10^6 M_{\odot}$ mass range from the Illustris cosmological simulation at $z>5$, considering both post-processed delays \citep{Kelley2017MassiveEnvironments} and no delays. D19 predicts merger rates of at most 0.02 yr$^{-1}$ for heavy seeds at $z > 5$ in their model assuming instantaneous merging and a LW background of $30\, J_{21}$. Finally, we compare with the recent work from \citet{Liu2024GravitationalMergers}, which uses an analytical model to reproduce JWST LFs for AGNs at $z=5$ using a mixture of heavy and light BH seeds. They report merger rates for $z>5$ of 0.1 yr$^{-1}$ for heavy seeds, and $32$ yr$^{-1}$ for the mixture of seeds.

We now calculate the merger rate as a function of redshifted merged BH mass $M_{z, {\rm BH}} = (1+z) M_{\rm BH}$. Here, $M_{\rm BH}$ corresponds to the total mass of the BH merger and $z$ the redshift at which the merger event occurred.

\begin{align}
    \frac{d^2 N}{dM_{z,{\rm BH}}\, dt} = 4 \pi c \sum_{N \in \Delta M_{z, {\rm BH}}} n(M_{z,{\rm BH}})\, \left( \frac{d_L(z)}{1+z} \right)^2 [{\rm yr}^{-1}]    
\end{align}

\begin{figure}
    \centering
    \includegraphics[width=\columnwidth]{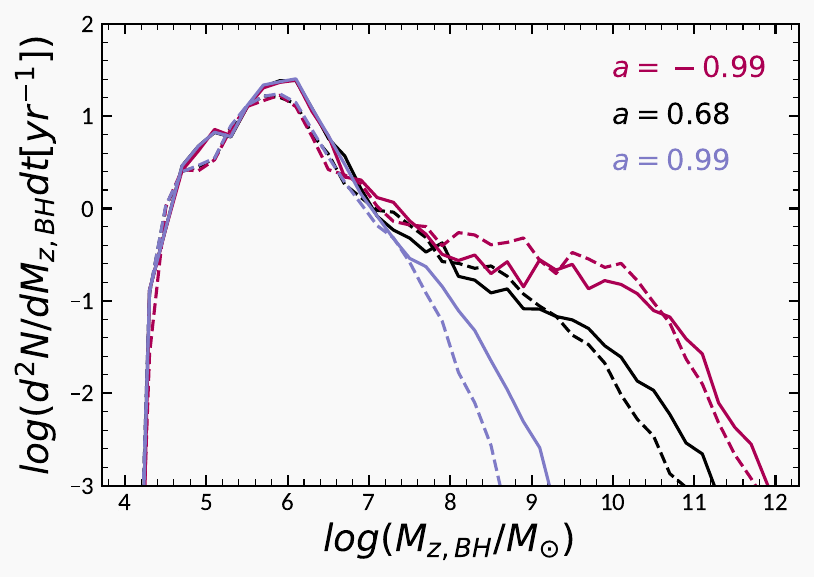}
    \caption{Merger rate events per year as a function of redshifted merged BH mass for all our models. Dashed lines correspond to the delayed models ({ \sc del}, del-sp099, del-sn099), and the solid to the instantaneous (fid, ins-sp099, ins-sn099)}
    \label{fig:merged_mass_rates_bhs}
\end{figure}

Where $n(M_{z,BH})$ is the number density associated to the redshifted merged BH mass. We show the redshifted observed merged BH mass in Fig \ref{fig:merged_mass_rates_bhs} for all models presented in Table \ref{models}. The dashed lines correspond to the delayed models, while the solid lines are for the instantaneous models. Each curve is colored according to its spin configuration. 

We start discussing the distribution that each model predicts on the observed redshifted BH mass shown in Fig. \ref{fig:merged_mass_rates_bhs}. For redshifted BH masses of $M_{z, {\rm BH}}\leq10^{6.7}M_{\odot}$ models with the same merging timescale, regardless of spin, exhibit overlapping distributions in merger rates. This agreement arises because these small BHs reside in small systems, where accretion is limited, making mergers the dominant growth channel. 
For $M_{z, {\rm BH}}>10^{6.3} M_{\odot}$ models begin to diverge. As expected, models with extreme prograde reach lighter redshifted masses at the high mass end, surpassed by the fiducial and the extreme retrograde. For a given spin, models with an instantaneous prescription populate towards heavier redshifted BH masses, compared to the corresponding delayed model. Moreover, the discrepancies between instantaneous and delayed models at a given spin become less pronounced towards more negative spins. For instance, at a fixed merger rate of $10^{-2}$ yr$^{-1}$, the redshifted BH mass reaches $M_{z, BH}/M_{\odot} \approx 10^{8.72} (10^{8.23})$ for ins-sp099 (del-sp099), $\approx 10^{10.52} (10^{10.09})$ for the fiducial ({\sc del}), and $\approx 10^{11.25} (10^{11.15})$ for ins-sn099 (del-sn099). Understandibly, accretion rates increases towards negative spins, accelereting BH mass growth through cosmic time, and partially compensating for the setback growth that delayed merging yields.

When comparing our fiducial model against the previous literature \citep{Sesana2007TheStream, Bonetti2019Post-NewtonianLISA, Barausse2020MassiveFeedback, Klein2016ScienceBinaries, Katz2020ProbingLISA}, we find that our distribution of merger rates as a function of redshifted BH mass show similarities among those who report relatively high merger rates for heavy redshifted BH mass \citep{Klein2016ScienceBinaries, Katz2020ProbingLISA, Barausse2020MassiveFeedback}, with a prominent peak around $M_{z, {\rm BH}} \sim 10^6 M_{\odot}$, which extends to $10^9 M_{\odot}$ for merger rates of $\approx$ 0.1 yr$^{-1}$. 
Nonetheless, it is not straightforward to draw a detailed comparison among these works, since at times they employ different signal to noise cuts corresponding to LISA's detectability, and consider merger rates down to redshift 0.

\section{Discussion and Conclusions}

\label{sec_conclusions}
In this work, we address the merger rates of BHs in the first Gyr of the Universe, in light of the overabundance of AGNs as reported from recent JWST discoveries \citep{Maiolino2023JADES.Mighty, Harikane2023JWST/NIRSpecProperties, Greene2023UNCOVERz5, Kokorev2024AFields, Matthee2023LittleSurveys, Akins2024COSMOS-Web:Assembly, Yue2023EIGERzgtrsim6}. We use{ \sc Delphi}, a Semi-Analytical model, that tracks the assembly of galaxies and BHs over a hierarchical evolution of dark matter halos. In our models, we seed heavy BH seeds ($10^3 - 10^5 M_{\odot}$) for each starting halo down to redshift $\approx$ 13, which accretes efficiently (at the Eddington limit) once the halo surpasses a critical threshold \citep{Bower2017TheEnd}. Heavy seeds have become, among others \citep{Lupi2024SizeJWST, Trinca2024EpisodicUniverse}, one of the leading models to explain the observed massive AGNs at high redshifts ($ z \approx 4-10$). In this work, we explore models with instantaneous and delayed merging (from dynamical friction using \citet{Lacey1993MergerFormation}) between galaxies and BHs, as well as different spin configurations: canonical ($a=0.67$), maximal prograde ($a=+0.99$) and maximal retrograde ($a=-0.99$) (see Table \ref{models}). We set the fiducial model as the one with instantaneous merging and canonical spin.
For the fiducial model, we explore the parameter space in {\sc Delphi} and select the parameters (see Table \ref{parameters}) that simultaneously best fit the UV luminosity function (UVLF), stellar mass function (SMF) for Lyman Break Galaxies (LBGs), and the Black Hole mass function (BHMF), and the bolometric LF of AGNs from JWST discoveries. These optimised parameters are then applied to the rest of the models explored in this work. 
Regarding the BH accretion parameters that best fit JWST-AGNs observables, these are 1$-$2 orders of magnitude greater than previous versions of {\sc delphi} intended to fit pre-JWST AGN observables \citep{Dayal2019EarlyEffects}. However, this comes with the caveat that the theoretical framework from this model and in \citet{Dayal2019EarlyEffects} are not equal, since we introduced the cold gas fractions and star formation efficiencies from {\sc Sphinx}$^{20}$, as in \citet{Mauerhofer2025SynergisingYears}.

The main remarks from this work are the following:

\begin{itemize}

    \item Under our theoretical framework, we find that the fiducial model agrees with the bolometric LF at redshift 6, while it underpredicts the number densities by $0.9-3.2$ dex at the bright end ($\geq 10^{45.5}$erg s$^{-1}$) for redshift 7, and overestimates by $0.22-1.6$ dex the number densities across all luminosities for redshift 5. Indeed, at z=7, only the ins-sn099 model is capable to reach such high number densities ($\approx 10^{-4}$cMpc$^{-3}$ dex$^{-1}$) in the bright end, as a result of their elevated BH accretion. Regarding the BHMF, the fiducial model tends to overpredict JWST estimations at redshift 5 and 6, except for those of \citet{Taylor2025Broad-LineDots}. However, by the nature of these estimations, they would not represent the entirety of the BH population. On one hand, \citet{Matthee2023LittleSurveys} and \citet{Taylor2025Broad-LineDots} only considers a sample of broad H$\alpha$ sources, while \citet{Kokorev2024AFields} assumes that all of their photometrically selected sample are at the Eddington limit.
    
    \item Including a dynamical timescale impacts the capacity to assemble massive BHs ($\geq 10^7 M_{\odot}$) at early times. After redshift $\approx$ 10, the merger rates for the delayed models become more frequent, and BH accretion (related to the spin configuration) becomes the main driver for BH growth. Interestingly, at redshift 5, the rapid BH accretion that the extreme retrograde models yield, virtually erases the impact of adding or not a delayed timescale (seen at Fig. 1).

    \item We find that the total number of merger rates for redshifts above 5 is 19.61 to 28.06 yr$^{-1}$ for the delayed and instantaneous model, respectively. These values are within the range found in previous work on heavy seed mergers \citep{Sesana2007TheStream, Klein2016ScienceBinaries, Barausse2020MassiveFeedback, Dayal2019TheAstronomy, Bonetti2019Post-NewtonianLISA, Katz2020ProbingLISA, Liu2024GravitationalMergers}. However, we caution that our delayed timescales are conservative and do not comprise the sub-kpc dynamics of interacting BHs. Regarding the merger rates as a function of redshifted mass, we report a distribution that favors larger masses, which peaks around $10^6 M_{\odot}$, and reaches the heaviest masses for models with the most negative spins.
    
\end{itemize}

There are still many uncertainties to address in both theory and observations. Regarding our models, we must bear in mind that by choosing conservative timescales for BH mergers, we report optimistic estimates of merger events per year. Indeed, in a model that comprises the full complexity of the sub-kpc evolution between BHs, such as the interactions from local gas \citep{Escala2005THEDISK, Dotti2006LaserDiscs}, neighboring stars \citep{Sesana2015ScatteringEnvironments, Vasiliev2015TheLimit, Pfister2019TheGalaxies}, and gravitational recoil \citep{Haiman2004CONSTRAINTSREDSHIFT, Izquierdo-Villalba2020FromEnvironments}, we expect a decrease in the number of BH merger events throughout cosmic time \citep{Barausse2020MassiveFeedback, Bonetti2019Post-NewtonianLISA}.
For instance, \citealt{Dosopoulou2017DynamicalProblem} finds that low-mass ratio binaries ($M_2/M_1  <  10^{-3}$) are expected to sink at longer timescales than one Hubble time. In the case of the fiducial, the mass ratio of the BH mergers starts with a median value of 0.31 at z$\sim$16, which stochastically decreases to 0.09 at z$\sim$5. Taking into account that the percentage of low-mass ratio binaries fluctuates between $7\times 10^{-5} \%$ at z$\sim$13 to 15.16$\%$ at z$\sim$5, we reiterate our optimism in the results. Certainly, given that we use the instantaneous model with the canonical spin as the fiducial, we are, in a sense, minimizing the accretion parameters for BHs to approach JWST observations. On the other hand, incorporating effects from third body interactions \citep{Bonetti2018Post-NewtonianSpace}, or BH accretion in satellite galaxies that are yet to merge with the main system (which in this work is set to zero), may counteract the slower BH mass assembly from mergers that a more complex model would yield. 

Additionally, our models do not account for the fact that BHs may experience different accretion histories depending on their location within the galaxy. In numerical studies it has been observed that off-nuclear BHs found in low-density environments would experience fewer accretion events in contrast to when nucleated \citep{Bellovary2021TheGalaxies}.

From our results, we also acknowledge the need for a more efficient mechanism to quench massive BHs $\geq 10^{9.5} M_{\odot}$, since, in our framework, BHs are allowed to grow as long as gas is available after star formation processes. Indeed, this work limits AGN feedback to a mechanical ejection of gas, with a strength that is fractionally equal for all systems ($f_w^{\rm BH}$ = 5$\times$10$^{-4}$). In turn, we have rare systems ($\leq 10^{-8}$cMpc$^{-3}$ dex$^{-1}$) harboring BHs of $\geq 10^{10} M_{\odot}$ which keep growing due to the large gas reservoir that their halos retain. A careful treatment of BH feedback may also avoid the overestimation we find in the bolometric LF at redshift 5. Additionally, a more diverse impact of BH feedback, would also help reproduce overmassive BHs as those reported in \citet{Maiolino2023JADES.Mighty}, \citet{Kocevski2024TheFields}. We can also try to reach these overmassive BHs by carefully implementing a physically motivated seeding mechanism for Direct Collapse (DC) BHs. In principle, heavy seeds from the DC channel take place in halos where early fragmentation onto stars must be avoided \citep{Agarwal2014}. This is possible due to the exposure to Lyman Werner radiation (11.2$-$13.6 eV) supplemented by first generation stars. This radiation photodissociates molecular Hydrogen (H$_2$) and creates Atomic Cooling Halos (ACHs), which are the birth sites for DCBHs. A heavy seed formed by the DC channel may be characterized by a slower assembly of their stellar component, which can explain these overmassive BH to stellar mass sources \citep{Visbal2018IdentifyingGalaxies}, compared to local relations. However, the seeding mechanism for heavy seeds from the DC channel reports extremely low number densities across redshifts ($\approx 10^{-2.88} - 10^{-9.2}$ cMpc$^{-3}$), too low to reproduce JWST observations in this work ($10^{-1.26}- 10^{-4.72}$ cMpc$^{-3}$). Nonetheless, we must be careful to understand how representative these overmassive BHs are to the underlying population, and whether the majority appear overmassive due to selection biases introduced from JWST instruments \citep{Li2024TipRelation}.

Finally, for a more accurate depiction of BH population through cosmic time, in future work we must also address the spin evolution as a function of their accretion and merger history. This work is now limited to bracketing the maximal and minimal growth of BHs given the two extreme spin scenarios ($a = \pm 0.99$).

Regarding observations, there is an ongoing debate on the true nature of LRDs sources, this being compact starburst galaxies \citep{Perez-Gonzalez2024WhatEdition}, dust-obscured AGNs \citep{Matthee2023LittleSurveys, Greene2023UNCOVERz5, Kokorev2024AFields} or a mixture of both \citep{Rinaldi2024NotDots}. Indeed, studies from photometrically selected LRDs find that SED fitting using Quasar or galaxy only templates can show a good fit to their data \citep{Akins2024COSMOS-Web:Assembly, Kokorev2024AFields}. Photometric selection of sources, may not only be prone to misidentification when it comes to its nature, but can also result in an interloper at lower redshifts \citep{Salvato2018TheRedshifts}. Regarding a plausible mixed nature, \citet{Rinaldi2024NotDots} found that 30$\%$ of their photometrically selected sample of LRDs show non-compact morphology in the UV rest-frame, where half of these show two distinctive sources while the other showcases asymmetries suggestive of ongoing or recent merger activity.
When spectroscopy is available and a broad balmer emission line is detected, studies can estimate the BH mass using local relations (z < 0.3) derived from reverberation mapping \citep{Bentz2013THENUCLEI}. It is debatable whether this relation stands for these high-$z$ sources (z > 4), since it is unknown whether the BLR is virialized when AGNs accrete close or above the Eddington limit \citep{Lupi2024SizeJWST,King2024TheQSOs}. Unfortunately, studies of reverberation mapping for high-$z$ sources are incredibly challenging due to the (1+z) lag increase, while other tracers for BH mass are minimal. Recently, \citet{Abuter2024AAgo} compared the dynamical measurement of the BH mass from bright quasars ($\log(L_{\rm Bol})=47.2-47.9$ erg s$^{-1}$) at $z \sim 2$, to BH single-epoch mass estimates from different broadened emission lines (CIV, H$\beta$, H$\alpha$), finding a mass overestimation of 0.43 to 1.2 dex compared to the dynamical method. There is a possibility that BH masses for high-$z$ AGNs may be overestimated from the single-epoch method. 
An AGN-only solution for LRDs is also questioned due to their mostly non-detected X-ray counterparts \citep{Ananna2024X-RayHoles, Maiolino2025iJWST/iSNe}. However, there are theoretical efforts that explain weak X-ray emission as a consequence of low spinning (a $\sim$ 0) and mildly-Super Eddington accretion 1.4 < $f_{Edd}$ < 4, which can lead to strong winds that artificially broaden the BLRs (boosting estimations for BH masses), as well as large covering fractions \citep{Pacucci2024MildlyDots}. Along this line, \citet{Taylor2025Broad-LineDots} found that stacking the spectra of sources with detected broad H$\alpha$ by those that enter the LRD criteria compared to those that do not, systematically report larger FWHMs (2188$_{-33}^{+33}$ km s$^{-1}$) compared to the latter (1768$_{-17}^{+18}$ km s$^{-1}$).  
Alternatively, \citet{Inayoshi2024BirthUniverse} suggested that BHs from LRDs may be maximally prograde spinning, hence the elevated radiative efficiency ($\epsilon$=0.42) would imply smaller BH masses. 
Given that the majority of the observables we aim to reproduce fall under the LRD criteria, showcasing an SED \citep{Akins2024COSMOS-Web:Assembly} that differs greatly from a typical AGN \citep{Shen2020The07}, a more detailed analysis for the UVLF will be shown in a follow-up paper. 
Finally, it is important to acknowledge that the estimated number densities for JWST-AGN could be biased due to cosmic variance (CV). The effect of CV worsens with increasing redshift and decreasing survey area. For high redshift surveys (z>4) \citet{Ucci2021AstraeusReionization} found that CV is negligible when considering a survey that covers an area $\geq 1000 arcmin^2$, which is not satisfied by any of the works that have reported number densities for JWST-AGNs. 


\section{Data availability}
The calibrated observables for AGNs and galaxies, as well as the merger rate events as a function of redshift and redshifted BH mass for all models explored, can be found \href{https://zenodo.org/records/17674477}{here}.

\begin{acknowledgements}
   We thank the anonymous referee for their suggestions on improving the presentation of this work. P. C\'aceres-Burgos acknowledges the support from the TITANS nucleo milenio NCN2023$\_$002. P. Lira acknowledges the support from the National Laboratory of High Performance Computing (NLHPC) Chile facilities, as well as the support from the FONDECYT 1201748. P. Dayal warmly acknowledges support from an NSERC discovery grant (RGPIN-2025-06182).

\end{acknowledgements}


\bibliographystyle{aa} 
\bibliography{references}

\begin{appendix}
    \section{Number densities of BH seeds}
    \label{Appendix:number_densities}
    
\begin{figure}
        \centering
      \includegraphics[width=\columnwidth]{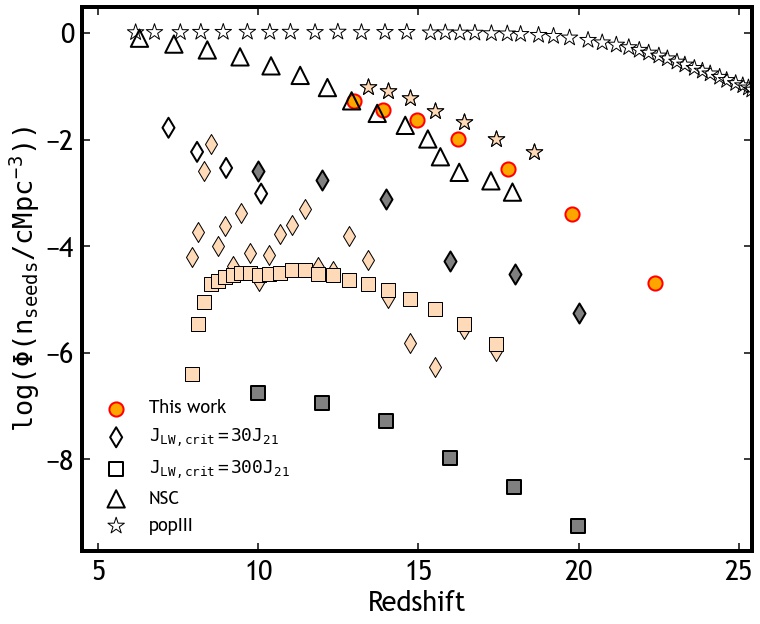}
        \caption{Number densities for different BH seeds by comoving Megaparsec. The orange circles are the case for seeding all pristine halos at redshift above $\gtrsim$ 13. The diamond shaped markers correspond to the DCBH scenario assuming the minimum Lyman-Werner radiation scenario, in white are \citet{Habouzit2016OnSeeds} and on gray \citet{Dijkstra2014Feedback-regulatedFormation}, the squares are the maximal Lyman-Werner threshold form \citet{Dijkstra2014Feedback-regulatedFormation}, the white stars are the semi-analytical calculation of NSC BH seed from \citet{Devecchi2012High-redshiftSeeds} assuming an average mass of $700M_{\odot}$.}
        \label{number_densities}
    \end{figure}

    In this work, we seed heavy BHs on all starting metal-free halos from our DM merger tree until redshift 12.8. Due to this decision is that the number density is subject to the resolution of the DM merger tree, and spans from $10^{-4.72}-10^{-1.26} cMpc^{-3}$ between redshifts 26 to 13. In Fig \ref{number_densities} we show our number densities (orange circles) compared to other numerical and semi-analytical simulations for light and heavy BH seeds. 
    Shown as squares and diamonds are the analytical and numerical predictions for physically motivated heavy BH seeds formed via the Direct Collapse method \citep{Dijkstra2014Feedback-regulatedFormation, Habouzit2016OnSeeds, Dayal2019TheAstronomy}. In squares (diamonds) are the predictions from assuming a critical LW radiation of $300 (30) J_{21}$. Using semi-analytical simulations \citet{Dijkstra2014Feedback-regulatedFormation} (shaded in gray) focuses on finding the abundance of halos that may form a DCBH, which satisfies $(i)$ a virial halo temperature ($T_{virial}$) above $10^4 K$, $(ii)$ low gas metallicity of $Z < Z_{crit} = 10^{-5 \pm 1} Z_{\odot}$ to prevent premature fragmentation, and $(iii)$ a LW exposure above a given threshold, able to photodissociate molecular hydrogen. Additionally, they explore the impact of metal pollution via galactic outflows (SNe and young stellar winds), as well as "genetic" enrichment from galaxy progenitors. We show their predictions from their fiducial models. Similarly, in hydrodynamical simulations, \citet{Habouzit2016OnSeeds} (empty diamonds) searches for eligible regions where DCBHs can form, requiring the criteria of $(i), (ii)$ and $(iii)$ and exploring different strengths of SNe feedback in three sets of numerical boxes with side lengths of 1, 10, and 142 cMpc. The data points we show for the latter work correspond to the number densities of halos in the 10cMpc with a prescription of weak SNe feedback. Similarly, 
    D19 (shaded orange) seeds heavy BHs in halos that satisfy $(i)$ $T_{virial} \geq 10^4 K$, $(ii)$ no metals $Z=0$, and $(iii)$ the $>0.5$ probability of the halo being radiated by a LW threshold above 30 or 300 $J_{crit}$. They also seed light BHs ($150M_{\odot}$) in metal-free starting halos until redshift 13 (orange stars), a prescription that aligns with the one used in this work for heavy seeds. However, we note that the number density from both our prescriptions of light seeds in D19 and our heavy seeds do not coincide as they should. This is because the HMF they use to calibrate they DM merger tree is a factor 2.6 greater than the one used in this work. Finally, we show the number density predictions from \citet{Devecchi2012High-redshiftSeeds} for PopIII BHs (empty stars) and BHs from the runaway collisions of Nuclear stellar clusters (empty triangles). These number densities are calculated using a semi-analytical prescription, that accounts for the metal enrichment from first generation stars, and the subsequent transition towards PopII and PopI stars, while for the formation of BHs via NSCs are tracked using the Toomre parameter of instability in disc like structure. We retrieve the data points for NSCs assuming an average mass of 700 $M_{\odot}$ as in \citet{Latif2019FormationHoles}.
    By comparing against these different works, we note right away how the physically motivated number densities predictions from BHs via the DC mechanism are too rare in comparison with this work. Indeed, to achieve similar observables as reported by JWST-AGNs, it is needed a mechanism of heavy seed formation that is much more abundant than the DC model. Nonetheless, our required number densities have comparable values as those reported by the runaway collisions of NSCs, however, the mass range that the latter method provides ranges between $10^2 - 10^3 M_{\odot}$, barely overlapping with the mass range from our initial BH mass prescription ($10^3-10^5 M_{\odot}$). Indeed, under the assumption that all reported JWST-AGN observations are accurate, and that these come from heavy seeds, it will be necessary to explore theoretical models that can allow for these to be more common.

    \section{Choosing the best fitting parameters}
    \label{best_fit_observables}
    In this section, we present the references used (Table \ref{table_param_Explore}) to determine the optimal parameters of our fiducial model (see Section \ref{sec:finding_params}), the values explored in our analysis, as well as the UVLF (Fig. \ref{fig:uvlf}) and SMF (Fig. \ref{fig:smf}) for our galaxy observables. 

    We note that for the UVLF of Lyman Break Galaxies (LBGs) (Fig \ref{fig:uvlf}), the fiducial agrees with observations for most redshifts except for those above z$\geq$11, and at z$\sim$5. At z$\sim$5, we see that galaxies with $M_{\rm UV}<$-20.5 underestimate the number densities found in observations by approximately 0.4 dex. We argue that this comes as a result of the elevated BH accretion and feedback found in massive galaxies, which compete with the gas mass budget for star formation. Regarding our predictions for the UVLF at z$\geq$11, we find a systematic underestimation of the number densities for bright galaxies with $M_{\rm UV} \lesssim$-18. In order to boost galaxy luminosities, previous work from  \citealt{Mauerhofer2025SynergisingYears} explored that by modifying the IMF as a function of redshift and metallicity, or the star formation efficiencies as a function of redshift, can result in a better fit to these extreme redshift datasets.
    
    \label{Appendix:best_fit_params}

    \begin{table}[]
    \caption{Grid of (free) parameters explored in this work}
\begin{tabularx}{\columnwidth}{|>{\raggedright\arraybackslash}p{0.42\columnwidth}|>{\raggedright\arraybackslash}X|}

\hline
Parameter                              & Explored values                                            \\ \hline
$f_{\rm Edd}(M_h<M_h^{\rm crit})$ &
  $\{ 10^{-5},  5 \times 10^{-5}, 10^{-4}, 5 \times10^{-4}, 10^{-3}, 5 \times 10^{-3}, 10^{-2}, 5\times 10^{-2}, 0.1, 0.5, 1 \}$ \\ \hline
$f_{\rm Edd}(M_h \geq M_h^{\rm crit})$ & 1                                                          \\ \hline
$f_{\rm acc}^{\rm BH}(M_h<M_h^{\rm crit})$ &
  $\{ 10^{-5}, 5 \times 10^{-5}, 10^{-4}, 5 \times 10^{-4}, 10^{-3}, 5 \times 10^{-3}\}$ \\ \hline
$f_{\rm acc}^{\rm BH}(M_h\geq M_h^{\rm crit})$ &
  $\{10^{-4}, 5 \times10^{-4}, 10^{-3}, 5 \times 10^{-3}, 10^{-2}, 5\times 10^{-2}, 0.1, 0.5, 1 \}$ \\ \hline
$f_w^{BH}$                             & $\{ 10^{-4}, 5 \times 10^{-4}, 10^{-3}, 5\times 10^{-3}\}$ \\ \hline
$f_{\star}$                            & $\{ 10^{-2}, 4 \times 10^{-2}, 6 \times 10^{-2}\}$         \\ \hline
\end{tabularx}

\label{table_param_Explore}
\end{table}

    \begin{table}[]
    \caption{References used for the chi square calculation of each observable.}
    \begin{tabularx}{\columnwidth}{|l|c|*{3}{>{\RaggedRight\arraybackslash}X|}}
    \hline
    Observable                                  & redshifts  & References used \\ \hline
    UVLF of LBGs              & 5-14         &    \citet{Bouwens2022Z2-9Turnover, Bouwens2021NewEfficiency,Bouwens2023EvolutionHUDF/XDF, Harikane2022GOLDRUSH.10,Harikane2023AEpoch,Harikane2024PureJWST/NIRSpec,Perez-Gonzalez2023LifeSurvey,Atek2015NewA2744,Atek2018TheUncertainties,Oesch2018TheMyr,Adams2022DiscoveryField,Donnan2022TheImaging, Donnan2024JWST15, Bowler2017UnveilingTelescope/i, Willott2024AFunction,Naidu2022TwoJWST,McLeod2024ThePrograms, Robertson2024EarliestBang, Whitler2025TheReionization, Leung2023NGDEEPImaging, Adams2024EPOCHS.Data,Casey2024COSMOS-Web:Assembly}    \\ \hline
    SMF of LBGs               & 5-10,12         &         \citet{Navarro-Carrera2024ConstraintsData, Song2016THEREDSHIFT, Duncan2014TheField, Bhatawdekar2019EvolutionFields, Stefanon2021GalaxyTime, Harvey2025EPOCHS.Observations}        \\ \hline
    BHMF of JWST AGNs                 & 5,6                  &       \citet{Kokorev2024AFields, Matthee2023LittleSurveys, Taylor2025Broad-LineDots}          \\ \hline
    bolometric LF of JWST AGNs & 5-8\tablefootmark{a}              &          \citet{Kokorev2024AFields, Akins2024COSMOS-Web:Assembly}       \\ \hline
    \end{tabularx}
    \tablefoot{ \tablefoottext{a}{When we use the $-$ sign on the second column of redshifts, we refer to all inclusive redshifts separated by bins of $\Delta z = 1$. For example,  5$-$8 = $\{5, 6, 7, 8\}$.}}
    \label{table:obser_chi2}
    \end{table}

    \begin{figure}
    \centering     
    \includegraphics[width=200px]{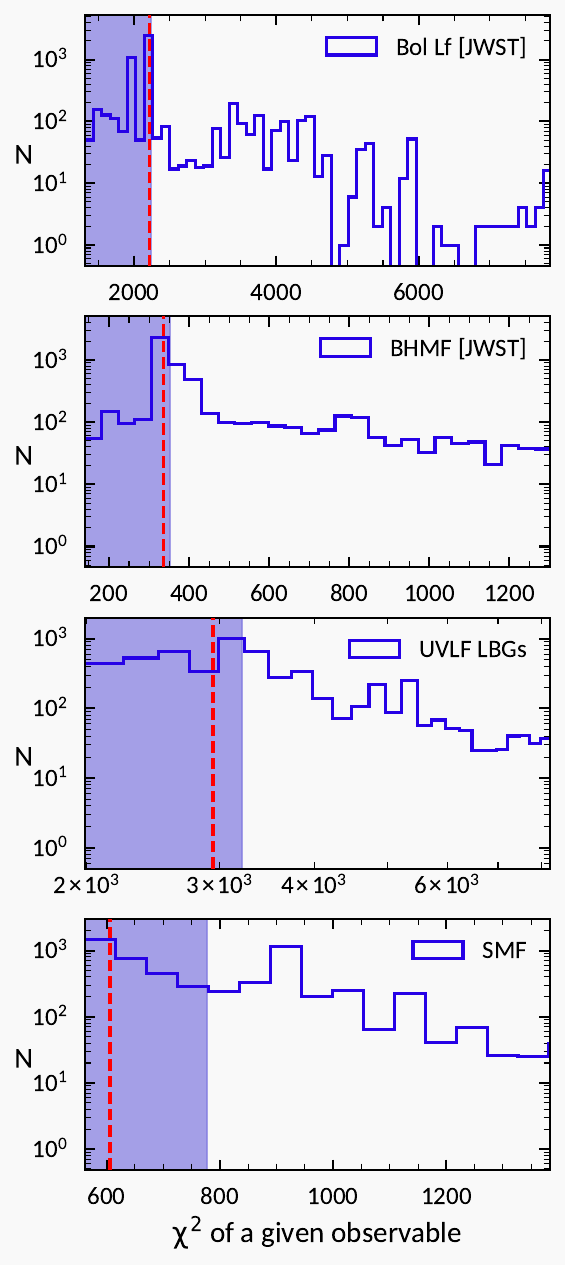}
    \caption{Chi square histogram from the parameter exploration for a given observable. From top to bottom: JWST bolometric LF, JWST BHMF, UVLF LBGs, and SMF. For each, we show in blue the models that are within the 50th percentile, and as a red vertical dashed line the model that satisfies our criteria of best overall fit}
    \label{fig:chi2}
    
\end{figure}

\begin{figure*}
    \centering
    \includegraphics[width=400pt]{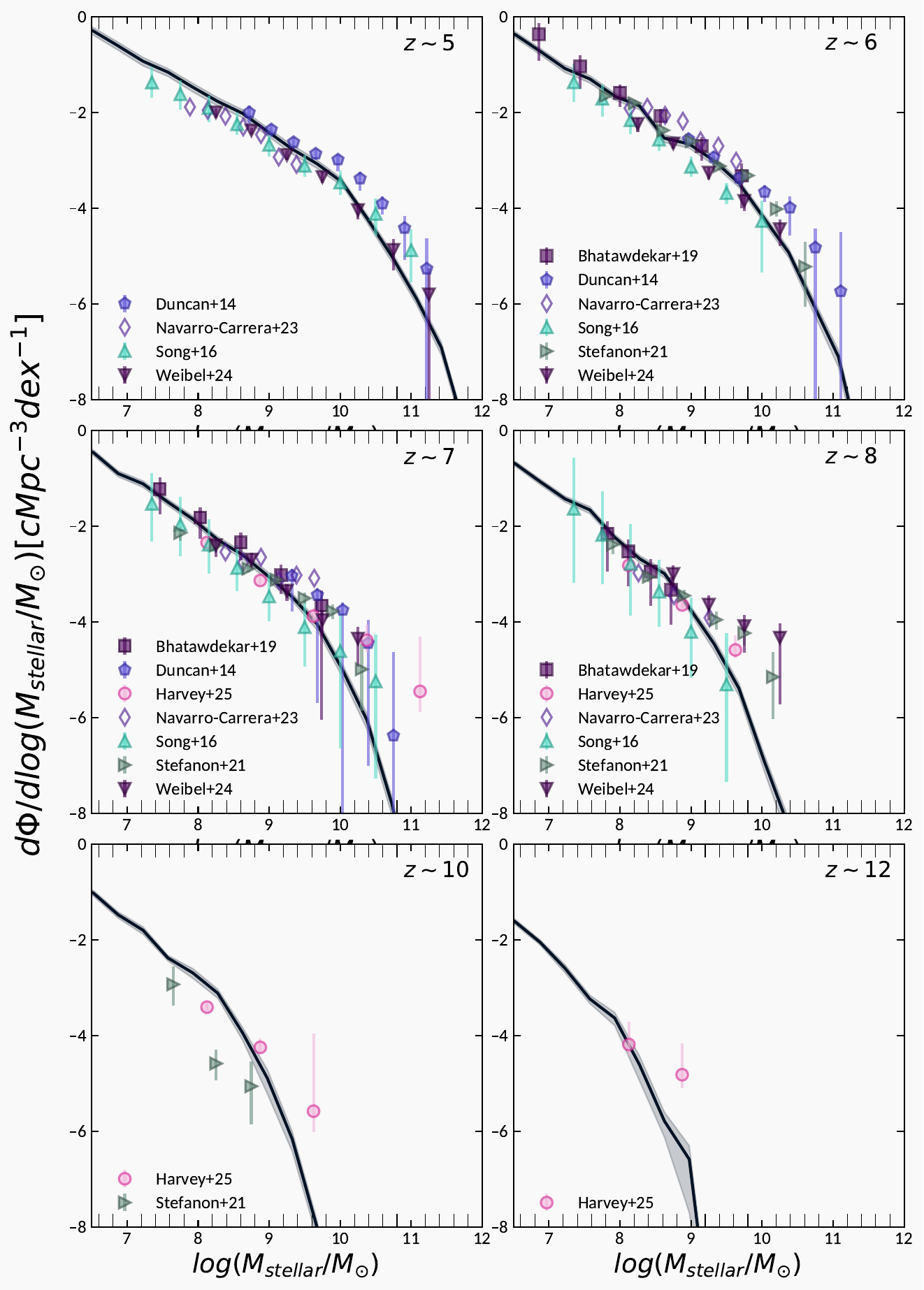}
    \caption{Stellar mass function of the fiducial model for redshifts 5, 6, 7, 8, 10 and 12. Observational data comes from \citealt{Duncan2014TheField,  Song2016THEREDSHIFT, Bhatawdekar2019EvolutionFields, Stefanon2021GalaxyTime, Navarro-Carrera2024ConstraintsData, Harvey2025EPOCHS.Observations}.}
    \label{fig:smf}
\end{figure*}

\begin{figure*}
    \centering
    \includegraphics[width=380pt]{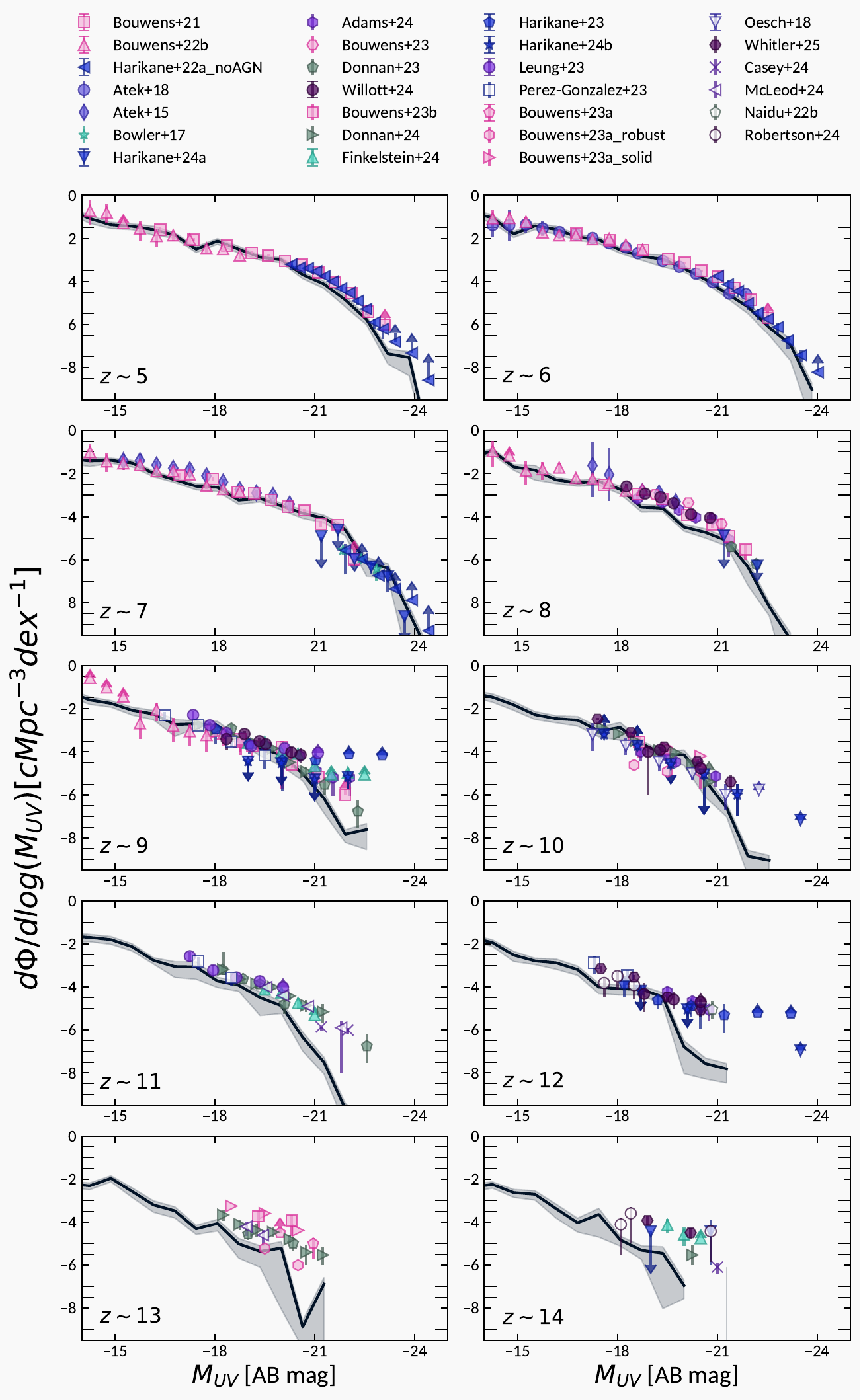}
    \caption{UV luminosity function of the fiducial model for redshifts 5 to 14. Observational data come from \citet{Bouwens2022Z2-9Turnover, Bouwens2021NewEfficiency,Bouwens2023EvolutionHUDF/XDF, Harikane2022GOLDRUSH.10,Harikane2023AEpoch,Harikane2024PureJWST/NIRSpec,Perez-Gonzalez2023LifeSurvey,Atek2015NewA2744,Atek2018TheUncertainties,Oesch2018TheMyr,Adams2022DiscoveryField,Donnan2022TheImaging, Donnan2024JWST15, Bowler2017UnveilingTelescope/i, Willott2024AFunction,Naidu2022TwoJWST, McLeod2024ThePrograms, Robertson2024EarliestBang, Whitler2025TheReionization, Leung2023NGDEEPImaging, Adams2024EPOCHS.Data,Casey2024COSMOS-Web:Assembly}.}
    \label{fig:uvlf}
\end{figure*}

\end{appendix}

\end{document}